\documentclass[twocolumn]{autart}    % Enable this line and disable the
\usepackage{amsmath,amssymb,amsfonts}
\usepackage{dsfont}
\usepackage{graphicx}
\usepackage{algorithm,algorithmic}
\usepackage{color}
\usepackage{cite}
\usepackage{subcaption}
\usepackage{epstopdf}
\usepackage{bm}
\usepackage{mleftright}
\usepackage{mathrsfs}
\usepackage{stmaryrd}
\usepackage{apacite}
\usepackage{booktabs}
\usepackage{graphicx}

\newtheorem{assumption}{Assumption}
\newtheorem{theorem}{Theorem}

\newtheorem{lemma}{Lemma}

\newtheorem{definition}{Definition}

\newtheorem{remark}{Remark}
\graphicspath{{./fig/}}
\begin{document}

\begin{frontmatter}

\title{
Distributed Nash Equilibrium Seeking under Quantization Communication  \thanksref{footnoteinfo}}

\thanks[footnoteinfo]{This work was supported in part by National Natural Science Foundation of China under Grant 61903027 and 72171172, by National Key R\&D Program of China under Grants 2018YFE0105000 and 2018YFB1305304, by Shanghai Municipal Science and Technology Major Project under Grant 2021SHZDZX0100, and by the Shanghai Municipal Commission of Science and Technology under Grants 1951113210 and 19511132101.}

\author[First]{Ziqin Chen}\ead{cxq0915@tongji.edu.cn},    % Add the
\author[First]{Ji Ma}\ead{maji@xmu.edu.cn},
\author[First]{Shu Liang}\ead{sliang@ustb.edu.cn},
and \author[First,ost]{Li Li}\ead{lili@tongji.edu.cn}
%\author[Third]{Ji Ma}\ead{maji@xmu.edu.cn}
%\author[HKU]{Sei Zhen Khong}\ead{szkhong@hku.hk},               % e-mail address
%\author[HKUST]{Li Qiu}\ead{eeqiu@ust.hk}  % (ead) as shown
\thanks[ost]{Corresponding author.}
\address[First]{School of Electronics and Information Engineering,
	Tongji University, Shanghai, P. R. China}
  % Please supply
%\address[SZ]{Independent researcher}             % full addresses
%\address[Second]{School of Electronics and Information Engineering,
%	Tongji University, Shanghai, P. R. China}
%\address[Third]{School of Aerospace Engineering,
	%Xiamen University, Xiamen, P. R. China}

\vspace{-1.5em}
                             % keyword list or with the
                                          % help of the Automatica
                                          % keyword wizard

\begin{abstract}
This paper investigates Nash equilibrium (NE) seeking problems for noncooperative games over multi-players networks with finite bandwidth communication. A distributed quantized algorithm is presented, which consists of local gradient play, distributed decision estimating, and adaptive quantization. Exponential convergence of the algorithm is established, and a relationship between the convergence rate and the bandwidth is quantitatively analyzed. Finally, a simulation of an energy consumption game is presented to validate the proposed results.
\end{abstract}
\end{frontmatter}

\section{Introduction}
%博弈在工程中的应用+分布式纳什均衡的发展
\vspace{-1em}
Game theory as a powerful tool for analyzing the interactions between rational decision-makers, has penetrated into various fields, including biology~\cite{biology}, economics~\cite{economy} and computer sciences~\cite{computersciences}. Nash equilibrium (NE), named after John Forbes Nash, Jr., is an important strategy profile of players in noncooperative games. Recently, advances in network optimization techniques have been applied to develop NE seeking algorithms~\cite{NEseeking1,NEseeking2,NEseeking3,NEseeking4,NEseeking5,NEseeking6,yuwenwu}.
\vspace{-0.5em}%
Note that the above NE seeking algorithms mainly focused on infinite precision transmission. However, the communication bandwidth is limited in the actual network, such as underwater vehicles and low-cost unmanned aerial vehicles systems. Hence, each player should sample and quantize its real value into finite bits before transmitting it while receiving it from its neighbors. This quantized communication process overcomes the bandwidth constraints, significantly reduces storage consumption, and is suitable for solving the practical network problems~\cite{quantizationeffect1,quantizationeffect2}. In the existing quantization works~\cite{yuandeming,dynamicquantizer1,dynamicquantizer3,liuyaohua,dynamicquantizer4}, the following three problems were mainly concerned: i) How can it ensure convergence even with inexact iterations throughout the distributed quantized algorithm? ii) What is the required minimum bandwidth when convergence is obtained? iii) How does the bandwidth affect convergence rate? To answer these problems, a zooming-in quantization rule is used in~\cite{dynamicquantizer1}, which proved that merely three bits could obtain the optimal solution. After that, only one-bit transmission was required in~\cite{dynamicquantizer3}, which explicitly characterized the proposed algorithm's sub-linear convergence rate. Further, the work~\cite{dynamicquantizer4} guaranteed a linear convergence rate of the quantized gradient tracking algorithm.

Although the above three questions have been widely discussed in distributed quantized optimization problems, few answers for the distributed NE seeking problem. Primarily because the cost function of each player in distributed NE seeking problems depends on the actions of all players, while the cost function of each player in a distributed optimization only depends on the action of itself. Thus, the update of the action of each player is much more complex in distributed NE seeking problems, which further brings technical difficulties in the design of the adaptive quantization scheme that depends on the trajectories of the actions of the players. Hence, the quantization scheme in distributed optimization problems can not be directly extended to distributed NE seeking, which motivates our works. Notably, the literature~\cite{duibi} tried to answer these problems for the distributed NE seeking, but in which each player was required to broadcast their quantized actions to all other players. It is still a centralized method in essence.

We take a step from our previous works on distributed NE seeking~\cite{liangshu} and distributed quantized cooperative problems~\cite{maji,xiaoqin} toward distributed quantized NE seeking. The main contributions are as follows.

%The authors in \cite{dynamicquantizer1} implemented a distributed quantized algorithm with zooming-in rule to solve an convex optimization problem under limited bandwidth constraint, and proved that merely three quantization level numbers could obtain the optimal solution. The authors in \cite{quantizationrate} further summary the convergence rates for distributed optimal algorithm when the communication between the players are quantized, and theoretical analysis demonstrated that the convergence rates match it under perfect communication through utilizing the adaptive quantization.
%%%%%%-------------------------------------------------------%%%%%%%%%
%第三段：本文做了哪些工作：设计了分布式算法，从定性方面分析了算法的收敛性以及收敛速度，从定量方面讨论了算法所需要传递的B位，讨论B与收敛速度的关系
%In this article, we focus on distributed NE seeking problems with finite bandwidth constraints, in which quantized actions are transmitted between neighbors. We analyze the convergence of the proposed algorithm and further compute the minimum bandwidth to ensure that the algorithm is exponentially convergent. Moreover, we give the relation between bandwidth and convergence rates.
%%%%%%-------------------------------------------------------%%%%%%%%%
%第四段：贡献：1. 解决了博弈中你所提出的问题，指数速度找到NE
%2.与现有的算法相比，a. 算法设计，考虑量化通信
%b. 优化设计，NE，量化两层估计。
%c.  收敛性分析，难点。（混杂系统，细节）
%3.给出定量的分析'

1) This is the first work to reveal that a distributed quantized NE seeking algorithm achieves exponential convergence under any positive bandwidth.
\vspace{-0.5pt}

2) An affine inequality explicitly characterizes the relation between the convergence rate and bandwidth, which indicts linearly increased convergence rate would linearly increase the bandwidth requirement.
\vspace{-0.5pt}

3) Our work is an extension to the distributed NE seeking with infinite precision transmission~\cite{NEseeking1,NEseeking2,NEseeking3,NEseeking4,NEseeking5}. Further, the assumption on the Lipschitz condition of the augment game mapping is not required anymore.
	
4) Compared with the only distributed quantized NE seeking work~\cite{duibi}, the communication graph must be fully connected. Our algorithm is distributed, and each player only interacts the quantized information with its neighbors, not all other players.

The rest of the paper is organized as follows. In Section \ref{sec:pre}, the problem is formulated. In Section \ref{algorithmdesig}, we propose the distributed quantized NE seeking algorithm based on the designed adaptive quantization scheme. In Section \ref{mainresult}, the main results, including the convergence analysis and the quantitative analysis on bandwidth, are discussed. An energy consumption game example is presented in Section \ref{sec:example} and the conclusion is given in Section \ref{sec:conc}.

\vspace{-5pt}
\textit{Notation:}
Denote $\mathbb{R}^{n}$ as the $n$-dimensional Euclidean space. For $x\in{\mathbb{R}^{n}}$, denote the $2$-norm by $\|x\|$. $\text{col}\{x_{i}\}_{i\in{\mathcal{I}}}$ stacks the vector $x_{i}$ as a new column vector in the order  of the index set $\mathcal{I}$. For matrices $A$ and $B$, the Kronecker product is denoted as $A\otimes B$. Denote by $0_{n},~1_{n}\in{\mathbb{R}^{n}}$, and $I_{n}\in{\mathbb{R}^{n\times n}}$ the vectors of all zero and ones, and the identical matrix.
A function $J:\mathbb{R}^{n}\rightarrow \mathbb{R}$ is strictly convex if, for all $x,y\in \mathbb{R}^{n}$ and $x\neq y$, $J \big (t x\!+\!(1-\!t)y \big)\!<\! t J(x)\!+\!(1-t)J(y)$ with $t\in(0,1)$. A function $J(x):\mathbb{R}^{n}\!\rightarrow\!\mathbb{R}$ is radially unbounded on $\mathbb{R}^{n}$ if for every $x_{n}\!\in\!{\mathbb{R}^{n}}$ such that $\|x_{n}\|\!\rightarrow\!\infty$, we also have $J(x_{n})\!\rightarrow\!\infty$. For a differentiable function $J(x): \mathbb{R}^{n}\!\rightarrow\!\mathbb{R}$, its gradient $\nabla_{x} J(x)\!\!=\!\!\text{col}\{\frac{\partial J}{\partial x_i}\}_{i\in\{1,\cdots,n\}}\!\in\!{\mathbb{R}^{n}}$. The minimum integer not smaller than $a\!\in\!{\mathbb{R}}$ is denoted as $\lceil\! a\! \rceil$.
%A function $f$ is strongly convex with parameter $m$ (or $m$-strongly convex) if the function $f(x)-\frac{m}{2}\|x\|^2_{2}$ is convex.
\vspace{-1em}
\section{Problem statement}\label{sec:pre}
\vspace{-0.5em}
Consider the noncooperative game  $G=\{\mathcal{V},J_{i},x_i\}$, where $\mathcal{V}=\{1,\cdots,N\}$ is the set of players involved in the game. A variable $x_{i}\in \mathbb{R}^{n_{i}}$ is the action of player $i\in{\mathcal{V}}$. A differentiable function $J_{i}(x_{i},x_{-i})\in \mathbb{R}$ is the local cost function of each player $i\in{\mathcal{V}}$, where  $x_{i}\in{\mathbb{R}^{n_{i}}}$ is its own action and $x_{-i}\in{\mathbb{R}^{n-n_{i}}}$ for $n=\sum_{i=1}^Nn_{i}$ denotes all players' actions except player $i$.

The aim of the NE seeking is that each selfish player $i$ obtains $x_{-i}$ through communication for minimizing its own cost function $J_i(x_{i},x_{-i}):\mathbb{R}^{n}\rightarrow \mathbb{R}$. The definition of the NE is given as follows.
\vspace{-0.5em}
\begin{definition}\label{def:causality}(Nash Equilibrium)
	Given a game $G\!=\!\{\mathcal{V},J_{i},x_{i}\}$, a vector of actions
$x^*\!=\!(x^*_{1},\!\cdots,\!x^*_{N})\!\in\!{\mathbb{R}^n}$ is a NE if $J_{i}(x^*_{i},\!x^*_{-i})\!\leq\! \inf_{x_{i}\in{\mathbb{R}^{n_{i}}}} J_{i}(x_{i},\!x^*_{-i}),\forall i\in{\mathcal{V}}$ holds.
\end{definition}

%In this paper, we focus on distributed NE seeking problems under limited bandwidth or capacity communication channels. It means that on one hand, each player only communicates with its neighbors rather than all players; on the other hand, it can only broadcast (receive) limited quantized information instead of obtaining accurate actions to (from) its neighbors.
%A sequence of edges $(i_{1},i_{2}),(i_{2},i_{3}),\cdots,(i_{k-1},i_{k})$ with $(i_{j-1},i_{j})\in{\mathcal{E}}$ for all $j\in\{2,\cdots,k\}$ is called a path from agent $i_{1}$ to $i_{k}$. If there always exists a path between any two different nodes of the graph, then the graph $\mathcal{G}$ is called connected.

We describe the information sharing between players as an undirected and connected graph $\mathcal{G}=(\mathcal{V},\mathcal{E})$, where $\mathcal{V}$ as the vertex set and $\mathcal{E}\subseteq \mathcal{V}\times \mathcal{V}$ as the edge set. Denote $\mathcal{N}_{i}\subseteq \mathcal{V}$ as the set of neighbors of player $i$.The adjacency matrix of the graph $\mathcal{G}$ is denoted as $\mathcal{A}=[a_{ij}]_{N\times N}$, with $a_{ij}>0$ if $(i,j)\in \mathcal{E}$, and $a_{ij}=0$ otherwise.The corresponding Laplacian matrix is $L_{\mathcal{G}}=[l_{ij}]_{N\times N}$, with $l_{ij}=-a_{ij}$ if $i\neq j$, and $l_{ij}=\sum^N_{j\neq i}a_{ij}$ otherwise. For an undirected and connected graph $\mathcal{G}$, one has that $L_{\mathcal{G}}1_{N}=0_{N}$, $1^{T}_{N}L_{\mathcal{G}}=0^T_{N}$ and all eigenvalues of $L_{\mathcal{G}}$ are real numbers and could be arranged by an ascending order $0=\lambda_{1}<\cdots \leq \lambda_{N}$. To proceed, we further make the following technical assumptions.

\begin{assumption}\label{ap1}
	For every $i\!\in\!{\mathcal{V}}$, the local cost function $J_{i}\!(x_{i},\!x_{-i}\!)$ is continuously differentiable, strictly convex and radially unbounded in $x_{i}\!\in\!{\mathbb{R}^{n_{i}}}$ for any fixed $x_{-i}$.
\end{assumption}

Assumption \ref{ap1} was widely used in the existing related works such as Assumption 2 in \cite{NEseeking3} and Assumption 1 in \cite{NEseeking5}.
%Stack the $\nabla_{x_{i}}J_{i}(x_{i},x_{-i})\!=\!\partial J_{i}(x_{i},x_{-i})/\partial  x_{i}\in{\mathbb{R}^{n_{i}}}$ as
\begin{definition}
The game mapping $F(x):\mathbb{R}^{n}\rightarrow \mathbb{R}^{n}$ is defined as $F(x)=\text{col}\{\nabla_{i}J_{i}(x_{i},x_{-i})\}_{i\in{\mathcal{V}}}$.
\end{definition}
The following assumptions formulate the restricted strongly monotone and the Lipschitz continuity of the elements of the game mapping $F(x)$.
\begin{assumption}\label{ap2}
The game mapping $F(x)$ satisfies
\begin{itemize}
\vspace{-0.5em}\item $F(x)$ is $\mu$-strongly monotone with the constant $\mu>0$, that is, for any $x,y\in{\mathbb{R}^{n}}$
\vspace{-0.5em}
$$\langle F(x)-F(y),x-y\rangle\geq \mu\|x-y\|^2.\nonumber$$
\vspace{-1.5em}
\item For every $i\in{\mathcal{V}}$, the gradient $\nabla_{x_i}J_{i}(x_{i},x_{-i})$ is uniformly Lipschitz continuous in $x_{i}$, that is, there is some constants $\theta_{i}\!\!\geq\!\! 0$ such that for any fixed $x_{-i}\!\in\!{\mathbb{R}^{n-n_{i}}}$,
\vspace{-0.5em}
$$\|\nabla_{x_i}J_{i}(x_{i},x_{-i})-\nabla_{x_i}J_{i}(y_{i},x_{-i})\|\leq \theta_{i}\|x_{i}-y_{i}\|.\nonumber$$
Moreover, for every $i\in{\mathcal{V}}$ the gradient $\nabla_{x_i}J_{i}(x_{i},x_{-i})$ is uniformly Lipschitz continuous in $x_{-i}$, that is, there is some constants $\theta_{-i}\!\!\geq\!\!0$ such that for any fixed $x_{i}\in{\mathbb{R}^{n-n_{i}}}$,
\vspace{-0.5em}
$$\|\nabla_{x_{-i}}J_{i}(x_{i},x_{-i})\!-\!\nabla_{x_{-i}}J_{i}(x_{i},y_{-i})\|\!\leq\! \theta_{-i}\|x_{-i}\!-\!y_{-i}\|.\nonumber$$
\end{itemize}
\end{assumption}

Define $\theta\!=\!(\theta_{i}^2+\theta_{-i}^2)^{1/2}$. It follows from Assumption 3 in \cite{NEseeking3} and Assumption 2 in \cite{NEseeking5} that Assumptions 1-2 ensure the existence and uniqueness of the NE for the game $G$. %We also need the following assumption to guarantee the feasibility of quantization at the initial time $t=0$.}
\begin{assumption}\label{ap3}
The initial states of all players satisfy $\|x_{i}(0)\|_{\infty}\leq M$ for $i\in{\mathcal{V}}$ and $\|x^*\|_{\infty}\leq M'$.
\end{assumption}

\begin{remark}
It is worth pointing out that in existing works on distributed quantized consensus, the assumption on the initial state, that is, $\|x_{i}(0)\|_{\infty}\leq M$ was required to estimate
the upper bound of tracking errors, see Assumption 2 in~\cite{youkeyou} and  Assumption 3 in~\cite{maji}. These errors guide the design of the scaling function to
avoid the saturation of quantizers at the initial time. However, in the game context, the upper bound of tracking errors is related to both $x_{i}(t)$ and $x^*$. Hence, $\|x^*\|_{\infty}\!\leq\! M'$ is also needed.
\end{remark}
\vspace{-1em}
\section{Algorithm design}\label{algorithmdesig}
\vspace{-0.8em}
In the distributed framework, each player has no access to the exact action of all other players, and it only receives a fixed number of bits from its neighbors. Frequently that means each player $i,i\in{\mathcal{V}}$ needs to estimate all other players' actions. We denote this estimated action as $\boldsymbol{x}^{i}\!=\!(x^{i}_{1};\cdots;x^{i}_{N})\!\in\!{\mathbb{R}^{n}}$, where $x^{i}_{i}$ is actual actions of player $i$ and $x^{i}_{j}$ is an estimated action of player $j$.
\vspace{-0.5em}
\begin{figure}[H]
	\centering
	\includegraphics[height=4.5cm,width=8.5cm]{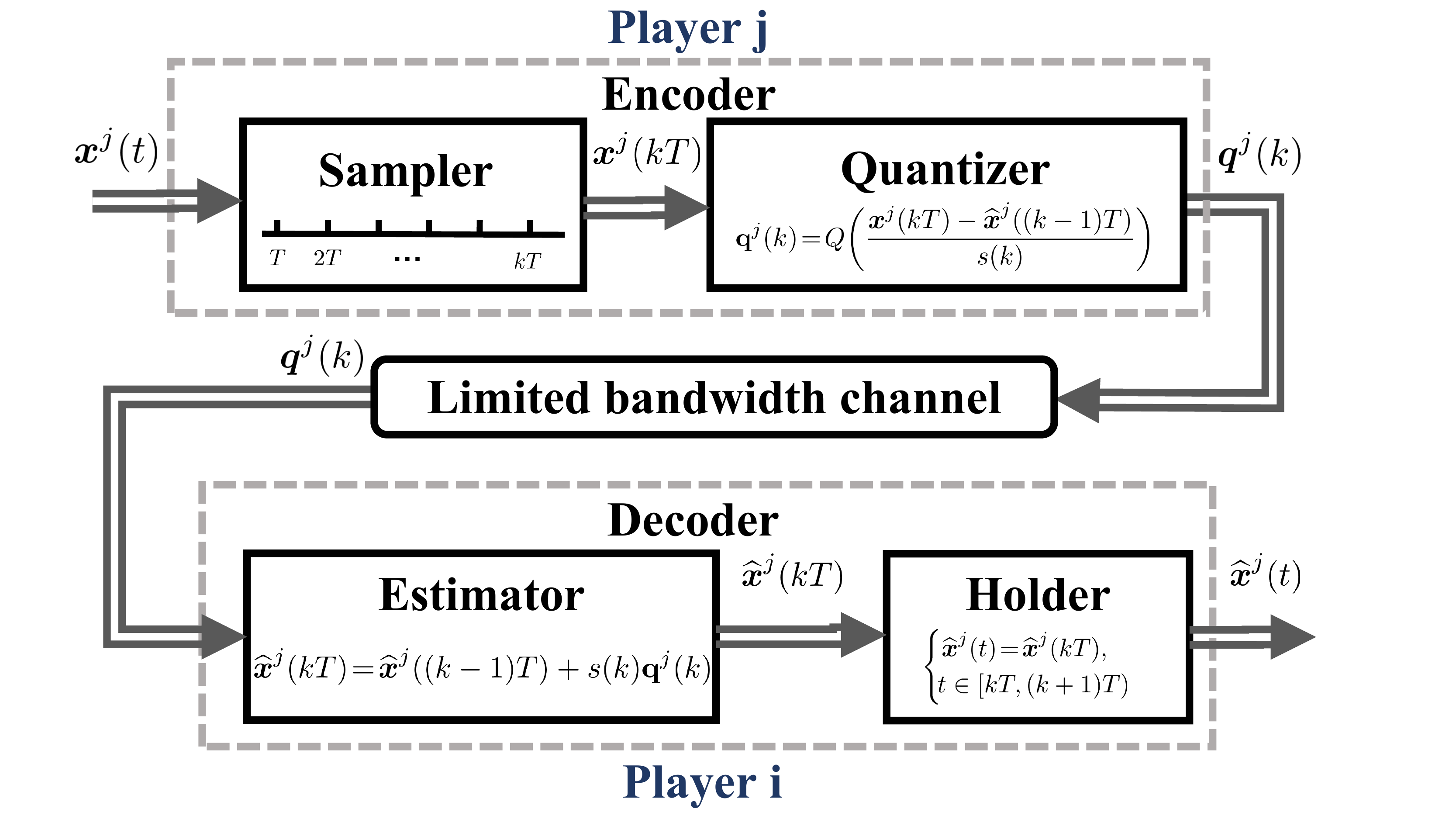}
	\caption{Communication Process}\label{fignetwork}
\end{figure}
\vspace{-0.5em}
Since the communication digital channels among players exist bandwidth constraints, each player $i$ interacts the quantized version of $\boldsymbol{x}^{j}$ with its neighbors $j,~j\!\in\!{\mathcal{N}_{i}}$. As shown in Fig. \ref{fignetwork}, quantized communication process for $(j,i)\in{\mathcal{E}}$ is summarized as the following two parts.

\vspace{-0.5em}
\begin{itemize}
\item \textbf{Quantized communication process for $(j,i)\in{\mathcal{E}}$.}

(i) Encoder: Player $j$ samples its own estimation $\boldsymbol{x}^{j}(t)$  at the fixed sampling time $kT,~k\!\in\!{\mathbb{N}}$, then it encodes $\boldsymbol{x}^{j}(kT)$ as the quantized message $\mathbf{q}^{j}(k)$ with a uniform quantizer as follows,
\vspace{-0.5em}
\begin{eqnarray}
	&&\mathbf{q}^{j}(0)\!=\!Q\left (\frac{\boldsymbol{x}^{j}(0)}{s(0)}\right )\nonumber,\\
	&&\mathbf{q}^{j}(k)\!=\!Q \bigg(\frac{\boldsymbol{x}^{j}(kT)-\widehat{\boldsymbol{x}}^{j}((k-1)T)}{s(k)}\bigg),~k\in{\mathbb{N}^+},\nonumber
\end{eqnarray}
where the multi-quantizer $Q(\cdot)\!\!=\!\!1_{n}\otimes q(\cdot)\in\mathbb{R}^{n}$ with $n\text{log}_{2}(2L+1)$ bits is designed as follows,
\vspace{-0.5em}
\begin{eqnarray}\label{quanter}
			\centering
			q(x)=\begin{cases}
				0,&\! \!\text{if}~-\frac{1}{2}< x < \frac{1}{2},\\
				i,& \!\!\text{if}~\frac{2i-1}{2}\!\leq x \!< \frac{2i+1}{2},\!~i\!=\!1,\cdots,L,\\
				L,&\! \!\text{if}~x\geq\frac{2L+1}{2},\\
				-q(-x),&\!\! \text{if}~x\leq-\frac{1}{2}.
			\end{cases}
	\end{eqnarray}

Then, player $j$ broadcasts the quantized message $\mathbf{q}^{j}(k)$ to its neighbor $i$ at time $kT$.

(ii) Decoder: Player $i$ receives $\mathbf{q}^{j}(k)$ from player $j$, then estimates $\boldsymbol{x}^{j}(t)$ as $\widehat{\boldsymbol{x}}^{j}(t)$.  The decoder  is designed as follows,
\vspace{-0.5em}
\begin{eqnarray}
	&&\!\widehat{\boldsymbol{x}}^{j}(0)=s(0)\mathbf{q}^{j}(0),\\
	&&\!\widehat{\boldsymbol{x}}^{j}(kT)\!=\! \widehat{\boldsymbol{x}}^{j}((k-1)T)+{s(k)}\mathbf{q}^{j}(k),\\
	&&\widehat{\boldsymbol{x}}^{j}(t)\!=\!\widehat{\boldsymbol{x}}^{j}(kT),~\!kT\!\leq\! t\!<\!(k\!+\!1)T,~k\!\in\!{\mathbb{N}^+}.\label{decode}
\end{eqnarray}

\vspace{-0.5em}
\item \textbf{Quantized parameters design:}
	
	 (i) Select the sampling period $T>0$ satisfying
	\begin{flalign}
		(e^{\alpha\lambda_{N}T}-1) (e^{\gamma T}-1)\rho\varepsilon^{-1}\leq a_1<1,\label{a1}
	\end{flalign}
where $\gamma\!=\!{\varepsilon\nu\beta}/{4}$, $\alpha\!>\!\frac{\varepsilon}{\lambda_{2}}(\frac{\theta^2}{\mu}\!+\!\theta)$, $a_{1},\beta\!\in\!(0,1)$,
	$\nu\!=\!2\lambda_{\min}\bigg(\!\bigg{[}\begin{matrix}
		\frac{\mu}{N} & -\frac{\theta}{\sqrt{N}} \nonumber\\
		-\frac{\theta}{\sqrt{N}} & \frac{\alpha\lambda_{2}}{\varepsilon}-\theta
	\end{matrix}\bigg{]}\!\bigg)$, $\rho\!=\!\left( \frac{\theta_{\mathbf{F}}\varepsilon}{\alpha\lambda_N}\!+\!1 \right)\!\frac{8\alpha\lambda_N}{\nu^2\beta\varepsilon\sqrt{1-\beta}}$.

	(ii) Design the scaling function $s(k)$ as follows,
	\begin{flalign}
		s(k)=s(0)e^{-\gamma kT}\label{scaling},
	\end{flalign}
	where $s(0)\!=\!\frac{ \varepsilon\nu M_0}{\alpha\lambda_N}\!\sqrt{1\!-\!\beta} e^{\!-\gamma  T\!-\!\alpha\lambda_NT}$ and $M_0\triangleq M\!+\!M'$.

	(iii) Choose $L$ as a positive integer satisfying
	$$L\geqslant\max\left\{\frac{M_{0}}{s(0)},\left\lceil\frac{\sqrt{Nn}e^{\varepsilon\nu\beta T/4+\alpha\lambda_N T}}{2a_2}-\frac{1}{2}\right\rceil\right\},$$
	where $0<a_2<1-a_1$.
\end{itemize}

\begin{remark}
Notably, an exponentially decaying scaling function~\eqref{scaling} is used here for the exponential convergence of the quantization errors, which is important to the exponential convergence of the NE seeking algorithm. On the other hand, $s(k)$ should be large enough such that the quantizer keeps non-saturated. That is, the convergence rate of $s(k)$ cannot be faster than that of the tracking error $\widetilde{\boldsymbol{x}}^i(t)=\boldsymbol{x}^i(t)-x^*.$ Hence, the designed convergence rate $\gamma$ for $s(k)$ matches that of $\widetilde{\boldsymbol{x}}^i(kT)$, whose convergence rate will be proved as $\gamma$ in the following Theorem 1.
\end{remark}

By using $\widehat{\boldsymbol{x}}^{i}(t)$ and $\widehat{\boldsymbol{x}}^{j}(t)$,  player $i$ updates its estimated action as follows,
\vspace{-0.5em}
\begin{flalign}
	\dot{\boldsymbol{x}}^{i}(t)=\alpha\sum^N_{j=1}a_{ij}(\widehat{\boldsymbol{x}}^{j}(t)-\widehat{\boldsymbol{x}}^{i}(t))-R_{i}\nabla_{i}J_{i}(\boldsymbol{x}^{i}),\label{algorithm}
\end{flalign}
where $R_{i}\!\!=\![\mathbf{0}_{n_{1}\times n_{i}},\!\cdots,\varepsilon I_{n_{i}\!\times\! n_{i}},\cdots,\mathbf{0}_{n_{N}\times n_{i}}]^{T}\in {\mathbb{R}^{n\times n_{i}}}$.

The dynamic \eqref{algorithm} is developed from \cite{NEseeking3,NEseeking4}, which requires continuous communication and accurate message interaction. In \eqref{algorithm}, each player just exchanges the quantized information with its neighbors at the sampling instant. Thus, our approach significantly saves communication resources.

%Additionally, the quantization runs in discrete time but iesies updates in continuous time. This complex hybrid system makes dynamics analysis difficult.
\begin{remark}
Compared with quantized NE seeking literature \cite{duibi}, in which the algorithm as
$x_{k+1}^{i}=x_{k}^{i}+\mu_{k} \frac{\partial}{\partial x^{i}} U_{i}\left(x_{k}^{i}, \mathcal{D}_{k}\left(\boldsymbol{x}_{k}^{-i}\right)\right)$, where $\mathcal{D}_{k}\left(\boldsymbol{x}_{k}^{-i}\right)$ represents quantized actions received from all other players at time $k$. It means that the communication graph is assumed to be fully connected, in contrast, we need not this assumption anymore.
\end{remark}
\vspace{-1em}
\section{Main results}\label{mainresult}
\vspace{-0.5em}
We prove that the dynamic \eqref{algorithm} exponentially converges to a NE in Subsection \ref{convergenceanalysis} and then discuss quantitative properties on the required bandwidth in Subsection \ref{relationship}.
\vspace{-0.5em}
\subsection{Convergence analysis}\label{convergenceanalysis}
\vspace{-0.5em}
First, we present the following lemma to prove that the equilibrium of the dynamic \eqref{algorithm} is a NE.
\vspace{-0.5em}
\begin{lemma}
The equilibrium $x^{*}$ of the dynamic \eqref{algorithm} is a NE of game $G$.
\end{lemma}
\vspace{-0.5em}
The proof is similar to the proof of Lemma 4 in \cite{NEseeking3} and thus omitted here.
%To proceed, we define the augmented mapping with  $\boldsymbol{x}=\text{col}\{\boldsymbol{x}^{i}\}_{i\in{\mathcal{V}}}$ as
	\begin{definition}
 The augmented game mapping is defined as $\mathbf{F}(\boldsymbol{x})\!=\!\text{col}\{\nabla_{i}J_{i}(x^{i})\}_{i\in{\mathcal{V}}}: \mathbb{R}^{n^2}\rightarrow\mathbb{R}^{n}$.
	\end{definition}
\vspace{-0.5em}
The following lemma shows the Lipschitz continuity of the augmented mapping $\mathbf{F}(\boldsymbol{x})$.
	\begin{lemma}
Under Assumptions 1 and 2, the augmented mapping $\mathbf{F}(\boldsymbol{x})$ is $\theta$-Lipschitz continuous in $\boldsymbol{x}\in\mathbb{R}^{n^2}$.
\end{lemma}
\vspace{-0.8em}
\begin{pf}
Follows from Assumptions 1-2, for any $x,y\in{\mathbb{R}^{n}}$ such that $x_{i},y_{i}\in{\mathbb{R}^{n_{i}}}$ and $x_{-i},y_{-i}\in{\mathbb{R}^{n-n_{i}}}$, there is
		\begin{eqnarray}
			&&\|\nabla_{i}J_{i}(x_{i},x_{-i})-\nabla_{i}J_{i}(y_{i},y_{-i})\|\nonumber\\
			&&=\|\nabla_{i}J_{i}(x_{i},x_{-i})-\nabla_{i}J_{i}(y_{i},x_{-i})+\nabla_{i}J_{i}(y_{i},x_{-i})\nonumber\\
			&&~~-\nabla_{i}J_{i}(y_{i},y_{-i})\|\nonumber\\
			&&\leq\Big(\beta \theta_{i}^2\|x_{i}\!-\!y_{i}\|^2\!+\!\beta/(\beta-1)\theta_{-i}^2\|x_{-i}\!-\!y_{-i}\|_{2}^2\Big)^{\frac{1}{2}},\label{lemma11}
		\end{eqnarray}
		where $\beta>1$. Choose $\beta=1+\theta_{-i}^2/\theta_{i}^2$ to rewrite \eqref{lemma11} as
		\begin{eqnarray}
			\|\nabla_{i}J_{i}(x)-\nabla_{i}J_{i}(y)\|\leq(\theta_{i}^2+\theta_{-i}^2)^{1/2}\|x-y\|,
		\end{eqnarray}
		Due to the arbitrary of $x,y\in{\mathbb{R}^{n}}$, Lemma 2 holds.
	\end{pf}
\vspace{-1.5em}
Note that the Lipschitz continuity of $\mathbf{F}(x)$ was assumed in most distributed NE seeking works, see Assumption 4 in~\cite{NEseeking3} and Assumption 5 in~\cite{NEseeking5}. Lemma 2 indicts that using Assumptions 1-2, the Lipschitz continuity of $\mathbf{F}(x)$ can be yielded such that it need not be assumed in this work.

%{\color{blue} Next,we prove Lemma 3 via the Lyapunov method, in which the Lyapunov function with respect to a simple coordinate transformation.} Define
Define $\varPhi_{1}\!\!=\!\!\frac{1}{\sqrt{N}}1_{N}$ and $\varPhi_{2}\!\!\in\!\!\mathbb{R}^{N\times(N-1)}$ to construct a unitary matrix $\varPhi\!\!=\!\![\varPhi_{1},\varPhi_{2}]$ such that $\varPhi^{T}L_{\mathcal{G}}\varPhi=\text{diag}(0,\lambda_{2},\cdots,\lambda_{N})$. Observe that $\varPhi^T_{2}\varPhi_{2}=I_{N-1}$ and $\varPhi_{2}\varPhi^T_{2}=I_{N}-\varPhi_{1}\varPhi^T_{1}$. Further, define tracking errors as $\widetilde{\boldsymbol{x}}^i\!=\!\boldsymbol{x}^{i}\!-\!x^*$ and estimation  errors as $\boldsymbol{e}^{i}\!\triangleq\! {\boldsymbol{x}}^{i}\!-\!\widehat{\boldsymbol{x}}^{i}$. Stack the above vectors as $\boldsymbol{x}^*\!=\!1_{N}\otimes x^*$, $\widetilde{\boldsymbol{x}}\!=\!\text{col}\{\widetilde{\boldsymbol{x}}^i\}_{i\in{\mathcal{V}}}$ and $\boldsymbol{e}=\text{col}\{\boldsymbol{e}^i\}_{i\in{\mathcal{V}}}$, respectively. The coordinate transformation of $\widetilde{\boldsymbol{x}}$ and $\boldsymbol{e}$ is written as follows,
\vspace{-0.5em}
\begin{flalign}
	&\overline{\boldsymbol{x}}_{1}(t)\!\!=\!\!(\varPhi^T_{1}\!\otimes\! I_{n})\widetilde{\boldsymbol{x}}(t)\!\!\in\!{\mathbb{R}^{n}},~\overline{\boldsymbol{x}}_{2}(t)\!\!=\!\!(\varPhi^T_{2}\!\otimes\! I_{n})\widetilde{\boldsymbol{x}}(t)\!\!\in\!{\mathbb{R}^{(N\!-\!1)n}},\nonumber\\
	&\overline{\boldsymbol{e}}_{1}(t)\!\!=\!\!(\varPhi^{T}_1\! \otimes\! I_{n})\boldsymbol{e}(t)\!\!\in\!{\mathbb{R}^{n}},~\overline{\boldsymbol{e}}_{2}(t)\!\!=\!\!(\varPhi^T_{2}\!\otimes\! I_{n})\boldsymbol{e}(t)\!\!\in\!{\mathbb{R}^{(N\!-\!1)n}}.\nonumber
\end{flalign}
\vspace{-2em}

Next, we will prove the exponential convergence of the quantized NE seeking dynamic \eqref{algorithm}, to do so,
 we present the following lemma whose proof  is given in Appendix.
\begin{lemma}
Construct the Lypaunov function as follows,
\vspace{-1em}
$$V(\overline{\boldsymbol{x}})=\frac{1}{2}(\|\overline{\boldsymbol{x}}_{1}\|^2+\|\overline{\boldsymbol{x}}_{2}\|^2).$$
Under Assumptions 1-3, along with the dynamic \eqref{algorithm}, if the following three inequalities hold when $k\!=\!k_{1},~\forall k_{1}\!\!\in\!{\mathbb{N}},$
\begin{eqnarray}
	&&\!\|\boldsymbol{e}(kT)\|  \!\leq\! \frac{a_2\varepsilon\nu M_0\sqrt{(1-\beta)Nn}e^{-\alpha\lambda_NT}}{2\alpha\lambda_N} e^{-\gamma(k+1) T} \label{l1},\\
	&&\!\|\boldsymbol{e}(t)\|\!<\!\frac{\varepsilon\nu M_0\sqrt{(1\!-\!\beta)Nn}}{2\alpha\lambda_N} \!e^{\!-\gamma(\lfloor t/T\rfloor\!+\!1) T},~t\!\in\![0,kT), \label{l2}\\
	&&\!V(\overline{\boldsymbol{x}})  \!\leq\! (NnM_0^2/2)e^{-2\lfloor t/T\rfloor T},~t\!\in\![0,kT],~k\in{\mathbb{N}} ,\label{l3}
\end{eqnarray}
then \eqref{l1}-\eqref{l3} hold when $k=k_1+1$.
\end{lemma}
%{\color{blue}whose proof is placed in Appendix. Actually, \eqref{l3} implies that the Lyapuove function $V(\overline{\boldsymbol{x}})$ converges to zero at the exponential convergence rate. We give the following theorem to illustrate $\boldsymbol{x}^{i}$ converges to $x^*$ exponentially.}
\begin{theorem}
Given an undirected and connected graph $\mathcal{G}$, under Assumptions 1-3, the dynamic \eqref{algorithm} exponentially converges to a NE of game $G$.
\end{theorem}
\vspace{-1em}
\begin{pf}
We prove Theorem 1 via the principle of induction. When $k\!=\!0$, it follows from $s(0)\!=\!\varepsilon\nu M_0/(\alpha\lambda_N)\!\sqrt{1\!-\!\beta} e^{\!-\gamma T\!-\!\alpha\lambda_NT}$ and Assumption 3 that \eqref{l1}-\eqref{l3} hold. Using the conclusion of Lemma 3, if \eqref{l1}-\eqref{l3} hold when $k=k_{1}$, it follows that \eqref{l1}-\eqref{l3} hold when $k=k_{1}+1$. We conclude that \eqref{l1}-\eqref{l3} hold for any $k\in{\mathbb{N}}$. By \eqref{l3}, $\overline{\boldsymbol{x}}$ exponentially converges to zero, which implies that $\widetilde{\boldsymbol{x}}(t)$ exponentially converges to zero.  Then, based on Lemma 1, Theorem 1 holds.
\end{pf}
%\begin{remark}Recalling the definition of $\widetilde{\boldsymbol{x}}(t)$, it yields $\boldsymbol{x}^{i}$ reaches the equilibrium of \eqref{algorithm} exponentially.
%Based on the definition of $V$ and the conclusions of Theorem 1, we denote the exponential convergence rate as $\gamma$ and it is no less than $\frac{1}{4}\varepsilon\nu\beta$, where $\nu=\lambda_{\min}\bigg(\bigg{[}\begin{matrix}
%	\frac{\mu_{F}}{N} & -\frac{\theta}{\sqrt{N}} \nonumber\\
%	-\frac{\theta}{\sqrt{N}} & \frac{\alpha\lambda_{2}}{\varepsilon}-\theta
%\end{matrix}\bigg{]}\bigg).$ In the lower bound of convergence rate $\frac{1}{4}\varepsilon\nu\beta$, the parameters $\frac{\alpha}{\varepsilon}$ and $\beta<1$ are the positive constants, $\lambda_{2}$ is the minimum non-zero singular value of matrix $\mathcal{L}_{\mathcal{G}}$, $\theta$ is the largest Lipschitz constant of $F$ and $\mathbf{F}$, and $\mu_{F}$ is the strongly monotone constant of $F$ and $N$ is the numbers of players.
%\end{remark}
\vspace{-0.5em}
\vspace{-1em}
\subsection{Quantitative analysis on bandwidth}\label{relationship}
\vspace{-0.5em}
In this subsection, Theorem 2 gives the required minimum bandwidth to ensure the exponential convergence of the dynamic \eqref{algorithm}. Theorem 3 discusses the relation between the required communication bandwidth and the convergence rate. The bandwidth is defined as follows.
\begin{definition} The bandwidth between the communication process $(j,i)\in{\mathcal{E}}$ is defined as
\vspace{-0.5em}
\begin{equation}
\mathcal{B}=\max_{(j,i)\in\mathcal{E}}\{\lim_{t\rightarrow\infty}\sup_{t'}\{\frac{1}{ t'}\sum_{t^{ji}_k\leq{t'}}r_{ji}(k),~t'\geq{t}\}\}~bits/sec,\nonumber
\vspace{-0.5em}
\end{equation}
where $t^{ji}_k,~k\in\mathbb{N}$ are the sampling instants and $r_{ji}(k)$ is the bits required to be transmitted at $t^{ji}_k.$
\end{definition}
\vspace{-0.5em}
\begin{theorem}\label{theorem2}
Given an undirected and connected graph $\mathcal{G}$, under Assumptions 1-4, the dynamic \eqref{algorithm} exponentially converges to a NE under any positive bandwidth.
\end{theorem}
%the dynamic \eqref{algorithm} exponentially converges to a NE based on Theorem 1, thus, we analyze the conditions to ensure Theorem 1 holds.
\vspace{-2em}
\begin{pf}
Choose $\frac{\alpha}{\varepsilon}>\frac{1}{\lambda_{2}}(\frac{\theta^2}{\mu}+\theta)$. In this case, the parameters $\nu$ and $\rho$ are two constants. For any $T>0$,
\begin{eqnarray*}
	\lim_{\varepsilon\rightarrow 0}(e^{\alpha\lambda_{N}T}\!\!-\!1) (e^{\frac{\varepsilon\nu\beta}{4}T}\!\!-\!1)\rho\varepsilon^{-1}\!\! \!=\!\lim_{\varepsilon\rightarrow 0}\!{\alpha\lambda_{N}T^2 \varepsilon\nu\beta\rho}/{(4\varepsilon)}\!=\!0,
\end{eqnarray*}
thus, \eqref{a1} is satisfied. If $\varepsilon$ is chosen properly, then
$$\lim_{\varepsilon\rightarrow 0}{\sqrt{Nn}e^{\varepsilon\nu\beta T/4+\alpha\lambda_NT}}/{(2a_{2})}\!=\!{\sqrt{Nn}}/{(2a_{2})}.$$
In this case, the transmitted quantized information $\boldsymbol{q}^{i}(k)$ is represented by $n\log_{2}\left(\!2\left\lceil \max\left\{\!\frac{M_{0}}{s_{1}},\frac{\sqrt{Nn}}{2a_2}\!\right\}\right\rceil\!+\!1\!\right)$ bits at each $T$. Since $T$ could be chosen by any positive constant, the bandwidth $\mathcal{B}$ could be any positive constants.
\vspace{-0.5em}
\end{pf}
 \begin{remark}
By Shannon’s rate-distortion theory, if there is a distributed algorithm achieving exponential convergence with the rate $\gamma$, then the communication bandwidth $\mathcal{B}\!>\!\gamma\log_{2}e\!>\!0$. Particularly, Theorem \ref{theorem2} establishes a sufficient and necessary condition on the required bandwidth for the exponential convergence of the dynamic \eqref{algorithm}.
\end{remark}
\vspace{-0.5em}
Naturally, much bandwidth means relaxed communication constraints, which contributes to the fast convergence rate for the dynamic \eqref{algorithm}. We give an affine inequality in the following theorem to describe this fact.
\begin{theorem}
	Given an undirected and connected graph $\mathcal{G}$, under Assumptions 1-3, the convergence rate and the minimum bandwidth required in the dynamic \eqref{algorithm} satisfy
	\vspace{-0.5em}
	\begin{flalign}
	\mathcal{B}\!\leq\!c_{1}\gamma\!+\!c_{2},\label{the2}
	\end{flalign}
where $c_1, c_2>0$ are some constants independent of  $\mathcal{B}$ and $\gamma$.
\end{theorem}
\vspace{-2em}
\begin{pf}
We prove \eqref{the2} via computing the upper bound of the minimum bandwidth $\mathcal{B}_{0}$ for any given convergence rate $\gamma_{0}$. Choose $\alpha/\varepsilon=c_{0}$ and $\beta$ as two positive constants such that $\nu$ is a positive constant.
Then, the convergence rate $\gamma_{0}=\varepsilon_{0}\nu\beta/4$ is determined by $\varepsilon_{0}.$ Let the sampling instants $T_{0}=1/(b_1\gamma_{0}+b_2)$, where $b_1=\frac{4c_{0}\lambda_{N}/(\nu\beta)+1}{\ln (\rho_0)}$, $b_2=\frac{\rho_0\rho\nu\beta}{4a_1}$ and  $\rho_0>1$. Using $T_{0}\!\leq\!\min\{1/b_1\gamma_{0},1/b_2\}$, $e^{\alpha\lambda_{N}T_{0}}\!-\!1\!\leq\! e^{\alpha\lambda_{N}T_{0}}$ and $e^{\gamma_{0}T_{0}}\!-\!1\!\leq\! \gamma_{0}T_{0}e^{\gamma_{0}T_{0}}$, we observe the chosen $T_{0}$ satisfies
\vspace{-0.8em}
\begin{eqnarray}\label{uighrfg}
  (e^{\alpha\lambda_{N}T_{0}}\!-\!1) (e^{\gamma_{0}T_{0}}-1)\varepsilon^{-1}\!\leq\! e^{\alpha\lambda_{N}T_{0}+\gamma_{0}T_{0}}{\nu\beta T_{0}}/{4}\!<\!{a_1}/{\rho},\nonumber
\end{eqnarray}
which grantees that $T_{0}$ satisfies \eqref{a1}. Next, we estimate the number of quantization levels $L_{0}$ for computing $\mathcal{B}_{0}$. Recalling from the definition of the bandwidth, it could be computed via
\vspace{-0.8em}
\begin{equation}
\mathcal{B}=\lim_{n\rightarrow \infty}\frac{\sum_{k=0}^{n-1}r_{ji}(k)}{nT}=\lim_{n\rightarrow \infty}\frac{\sum_{k=1}^{n-1}r_{ji}(k)}{(n-1)T}.
\end{equation}
It implies that the choice of the quantization levels at the initial time has no effect on the value of the communication bandwidth. Hence, we only consider the case $L_{0}\!=\!\left \lceil\!\dfrac{\sqrt{Nn}e^{\gamma_{0}T_{0}\!+\!\alpha\lambda_{N}T_{0}}}{2a_{2}}\!-\!\frac{1}{2}\!\right \rceil$ and we obtain
\vspace{-0.5em}
\begin{flalign*}
\mathcal{B}_{0}&\!\!={\log_{2}(2L_{0}+1)}/{T_{0}}= {\log_{2}(\sqrt{Nn}/a_2e^{\gamma_{0}T_{0}\!+\!\alpha\lambda_N T_{0}})}/{T_{0}}\nonumber\\
&\!\!{=}\log_{2}e(1\!+\!{4c_{0}\lambda_N}/{\nu\beta}\!)\gamma_{0}\!+\!{\log_{2} (nN)/2\!\!-\!\log_{2} a_2}/{T_{0}}\!\nonumber\\
&\!\!\overset{\Delta}{=}c_{1}\gamma_{0}+c_{2}, \label{equation}
\end{flalign*}
Since the chosen of $\gamma_{0}$ is arbitrary, Theorem 3 holds.
\end{pf}
\vspace{-2em}

\begin{remark}
The problem of the minimum bandwidth for the fixed convergence rate $\gamma_{0}$ is complicated and still unsolved in the quantized control. In fact, for a given convergence rate $\gamma_{0}$, Theorem 3 provides an upper bound of the minimum bandwidth $\mathcal{B}_{0}=c_{1}\gamma_{0}+c_{2}$, which partially deals with this problem.
\end{remark}
\vspace{-1em}
\section{An example}\label{sec:example}
\vspace{-0.5em}
In this section, we utilize an energy consumption game for heating ventilation and air conditioning systems \cite{NEseeking2} to illustrate the effectiveness of our results. The cost function of player $i$ is modeled as
\vspace{-0.5em}
$$f_{i}(x)\!=\!\!a_{i}\|x_{i}\!-\!b_{i}\|^{2}\!+\!x_i^T\left(\!c\sum_{i=1}^{N} x_{i}\!+\!d\!\right),\!~i=1,\cdots,5,$$
where $x_{i}\in{\mathbb{R}}^3$, $a_{i}=0.96-0.5i$, $b_{i}\!=\!\![9;11;13]\!+\!4(i-1)1_{3}$, $c\!=\!0.001$,  and $d\!=\!\![10;12;14]$. Based on theoretical analysis, the unique Nash equilibrium is computed as
$ x^{*}
	\!=\![x^*_{1};\cdots;x^*_{5}]\!=\![3.7608;4.7165;5.6722;7.4709;8.3692;
9.2675;11.1473;\\11.9816;12.8159;14.7838;15.5462;
16.3086;18.3724;\\19.0535;19.7345]\in{\mathbb{R}^{15}}.$
 The initial estimation is set as  $\boldsymbol{x}^{i}(0)\!=\![x_{1}(0);\cdots;x_{5}(0)]+2(i-1)1_{15}\in{\mathbb{R}}^{15}$ with  $x_{i}(0)\!=\![\!-\!10\!+\!3i;-5\!-\!2i;10\!-\!2i]$. The communication graph is given in Fig.2.

\begin{figure}[H]
	\centering
	\includegraphics[height=1.8cm,width=6cm]{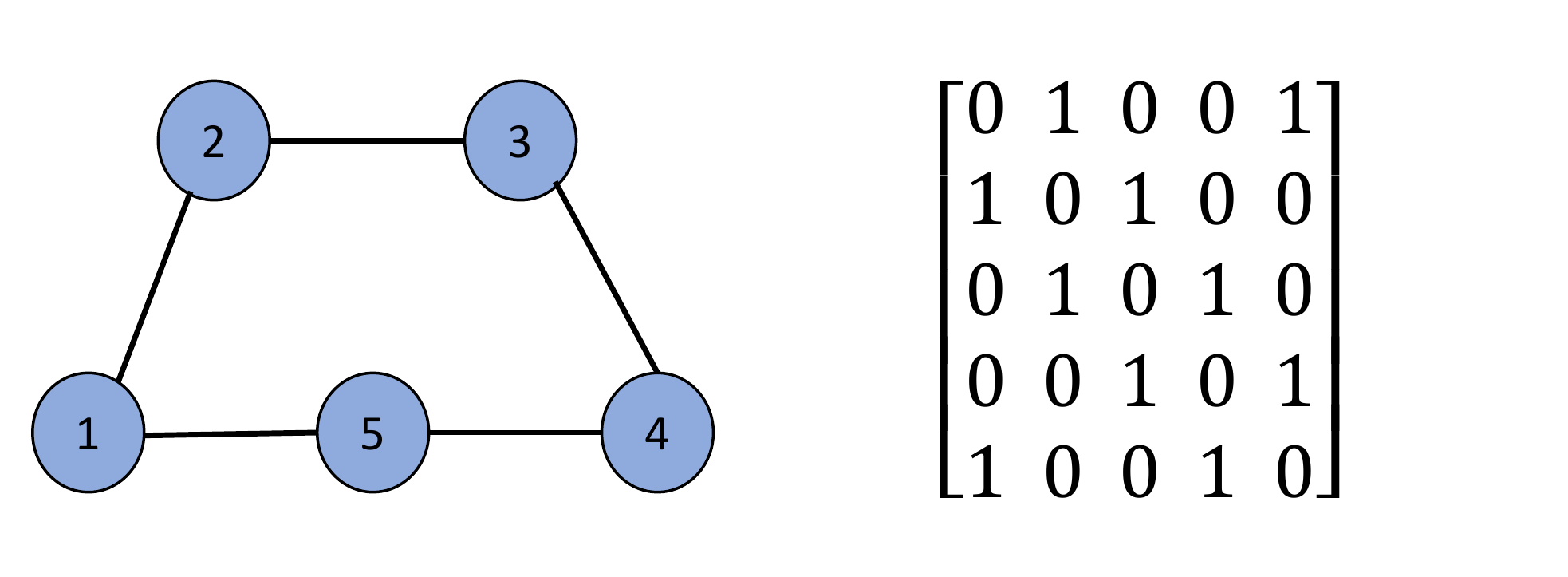}
	\caption{Communication graph.}
\end{figure}
\vspace{-1em}
The parameters of quantization scheme are chosen as: (a) the sampling period $T\!=\!0.1\text{sec}$; (b) the scaling function $s(k)\!=\!0.1e^{-0.1k}$; (c) the bandwidth $\mathcal{B}\!=\!270~ \text{bit/sec}$. 

We perform the proposed the dynamic \eqref{algorithm} with $\alpha=1$. Fig.3 compares theoretical NE and distributed estimated actions of all players for the three dimensions. It shows that the distributed estimates accurately track the theoretical NE [cf. Theorem 1].
\begin{figure}
	\centering
	\includegraphics[height=2.3cm,width=7cm]{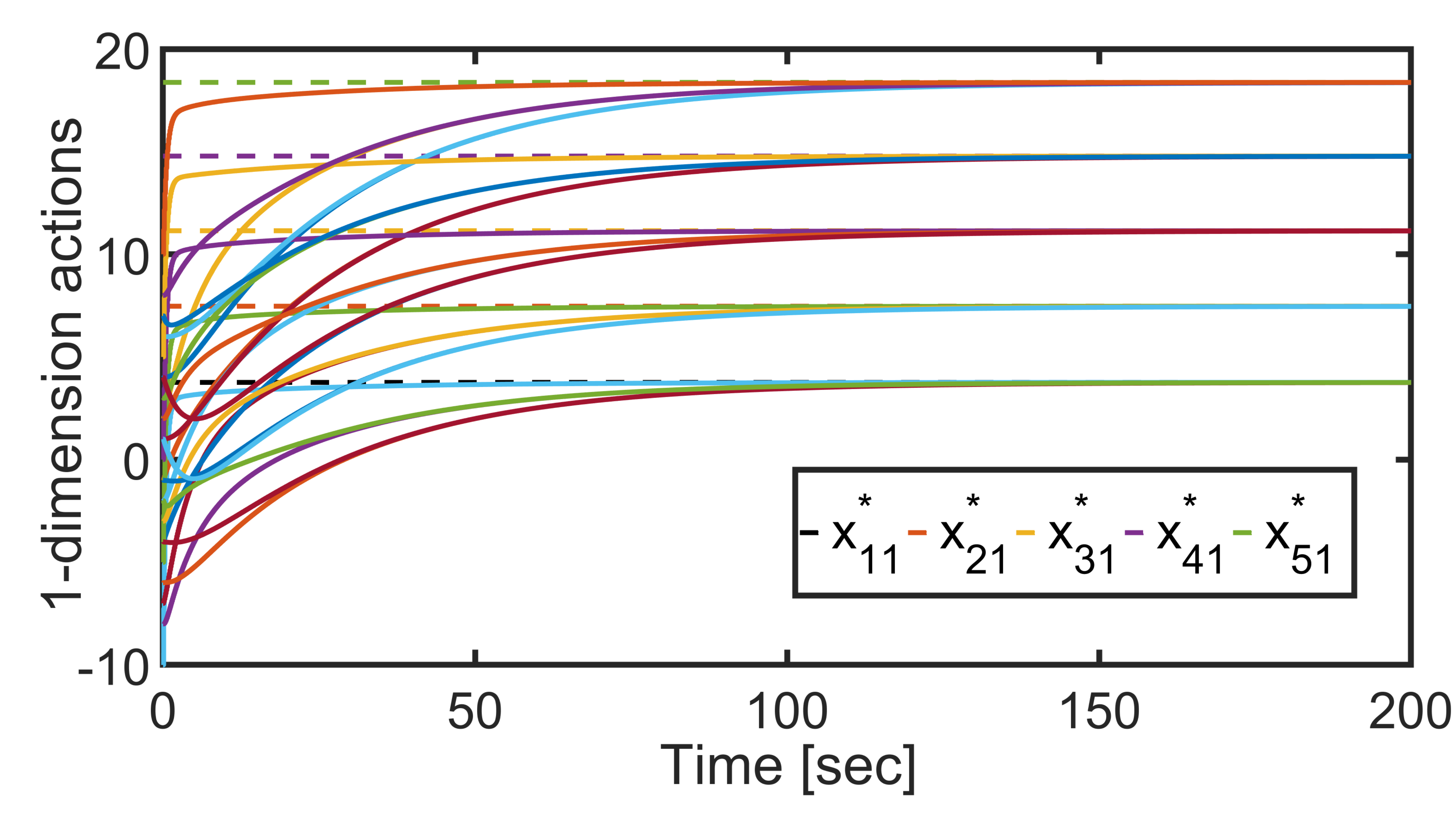}
	\includegraphics[height=2.3cm,width=7cm]{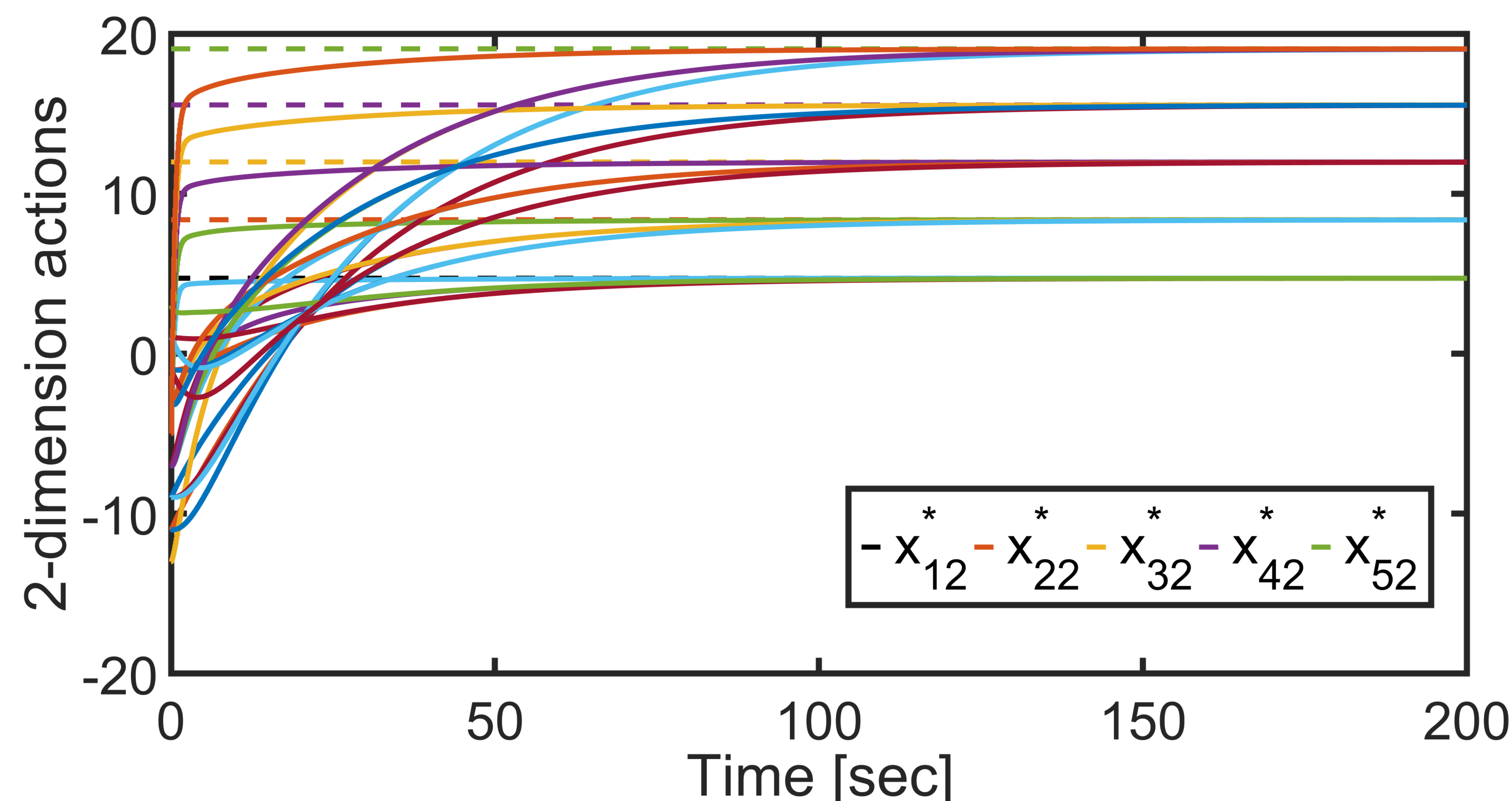}
	\includegraphics[height=2.3cm,width=7cm]{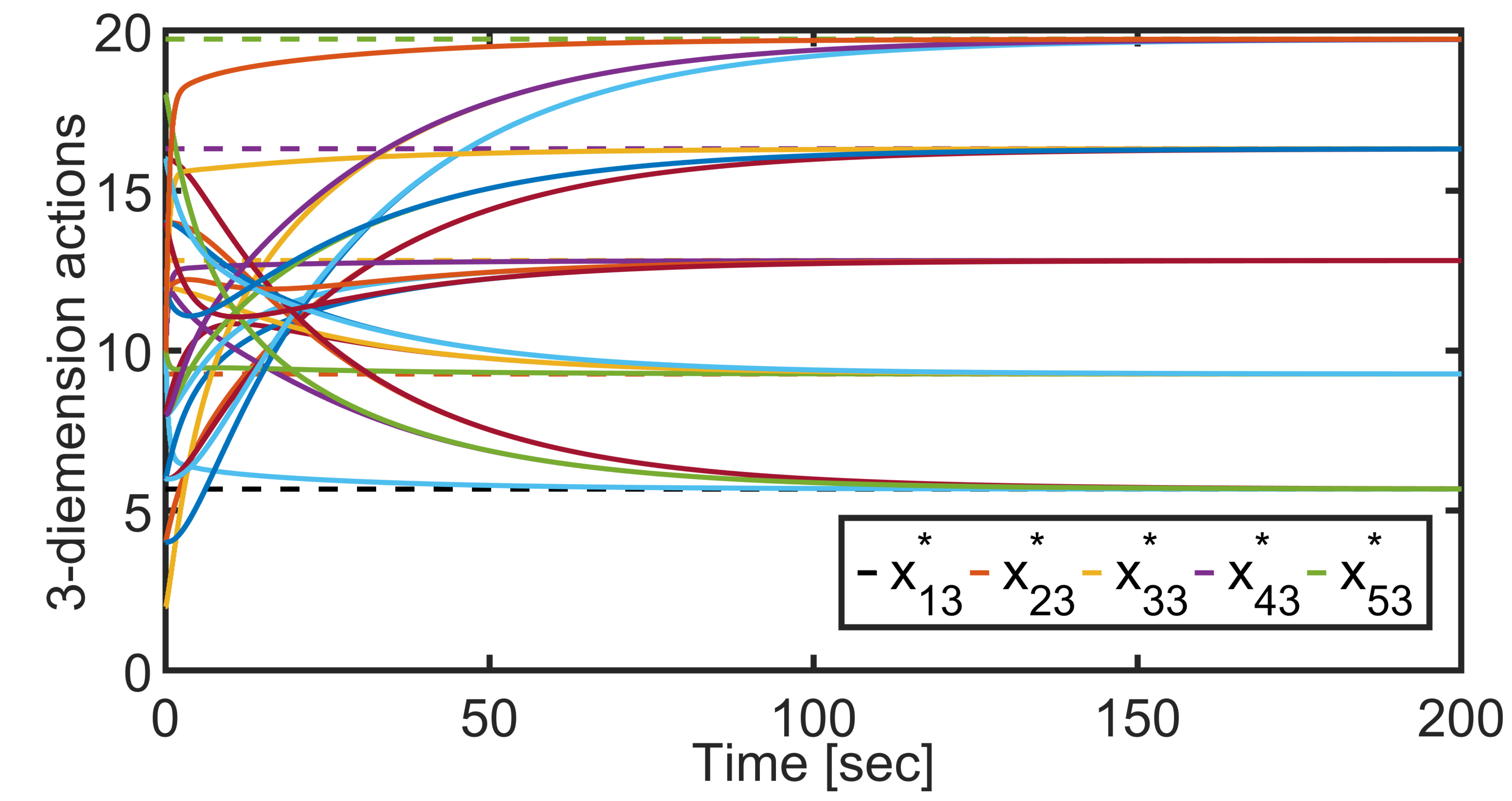}
	\caption{Three dimensions of the theoretical NE $x^*_{i}$ (dotted line), and dynamics $\boldsymbol{x}^{i}$  (solid line) for players $i,~i=1,\cdots,5$.}
\end{figure}

Fig. 4 compares the tracking errors $\|\boldsymbol{x}(t)\!-\!1_{5}\otimes x^*\|$ of our quantized algorithm with that of the existing distributed NE seeking algorithm presented in~\cite{NEseeking3,NEseeking4} under an ideal communication channel.  It shows that the quantized communication brings the difficulty to the NE seeking.
\begin{figure}
	\centering
	\includegraphics[height=2.3cm,width=7cm]{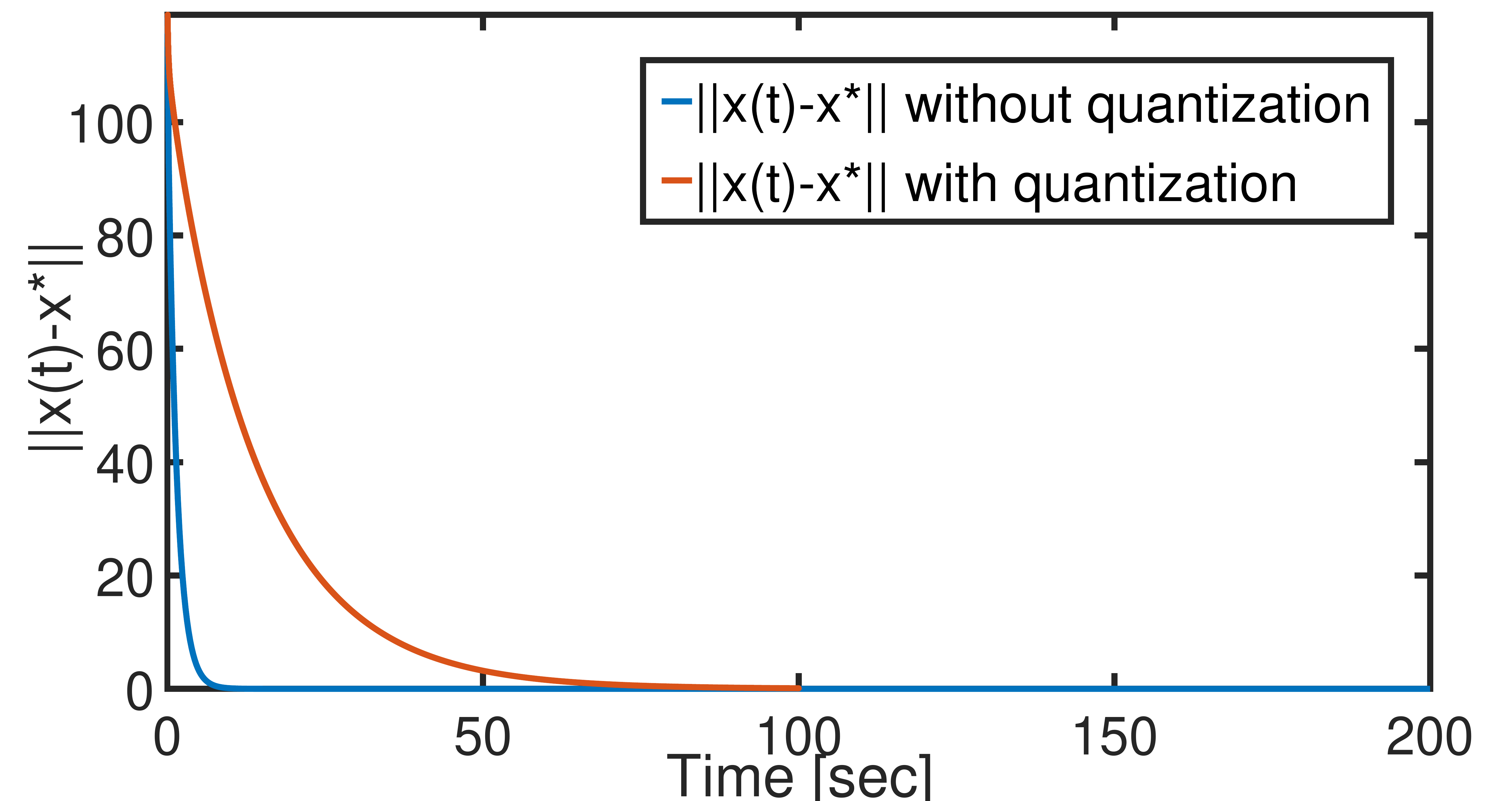}
	\vspace{-1em}
	\caption{The comparison of tracking errors between distributed NE seeking algorithm without quantization (blue line) and the dynamic \eqref{algorithm} (red line).}
\end{figure}

Fig. 5 shows simulation results for Theorem 3. It proves that for any given convergence rate $\gamma_{0}$, the actually required bandwidth $\mathcal{B}$ is less than the upper bound of the minimum bandwidth $\mathcal{B}_{0}=c_{0}\gamma_{0}+c_{1}$  [cf. Theorem 3].
\begin{figure}
	\centering
	\includegraphics[height=2.3cm,width=7cm]{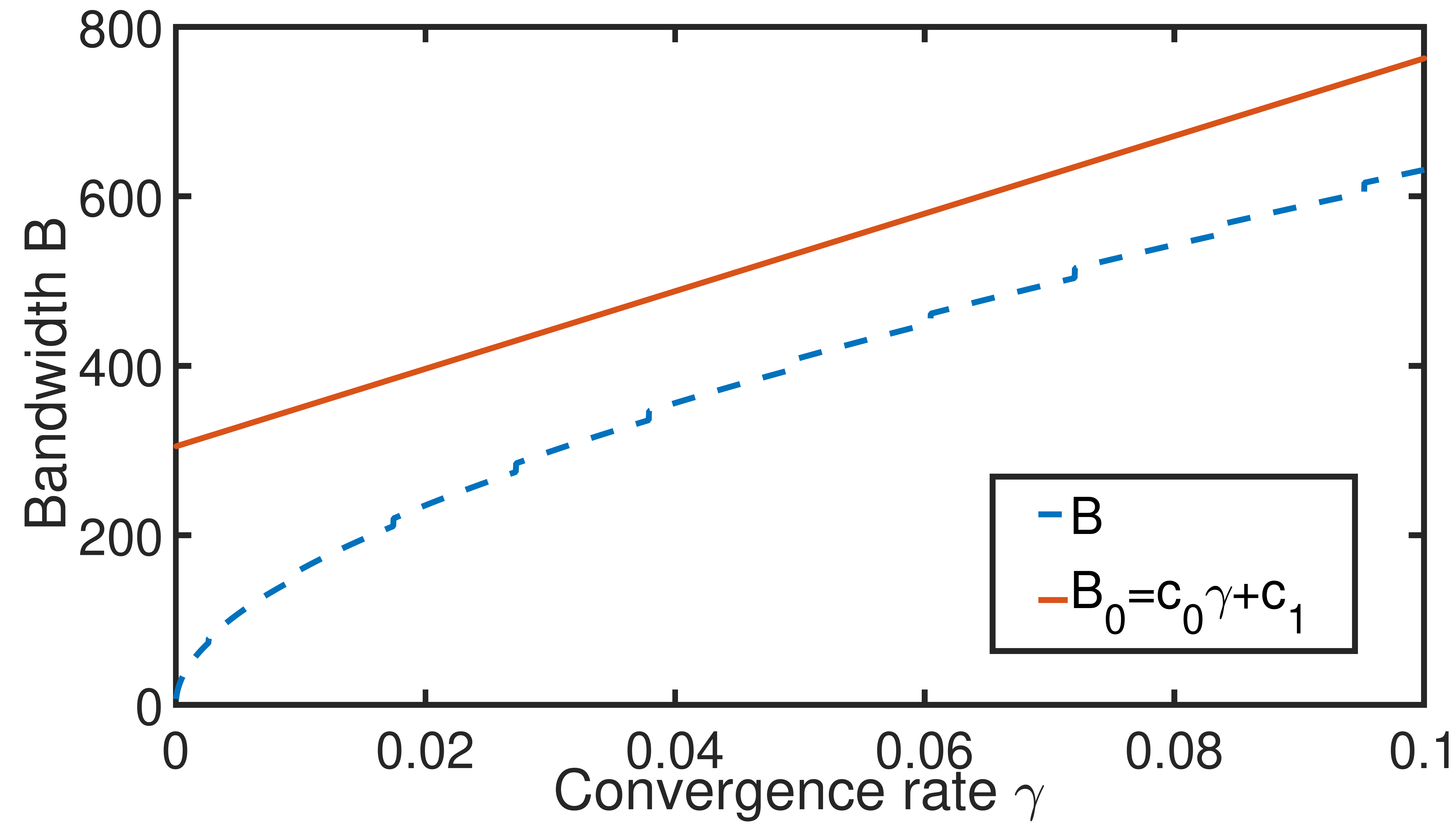}
	\caption{The upper bound of minimum bandwidth $\mathcal{B}_{0}$ (red line) and practical bandwidth $\mathcal{B}$ (blue line) for the given $\gamma$.}
\end{figure}
\vspace{-1em}
\section{Conclusions }\label{sec:conc}
\vspace{-1em}
We investigated the distributed NE seeking with finite bandwidth constraints among each pair of players.
To solve this problem, a distributed NE seeking algorithm with an adaptive quantization scheme was proposed. Theoretical and experimental results showed that for any bandwidth constraints, the proposed algorithm could achieve exponential convergence. In addition, an affine inequality was given to describe the relation between convergence rate and the required bandwidth.
\vspace{-0.8em}

\section{Appendix}\label{appendix2}
\vspace{-1em}
For notations simplicity, define
\begin{eqnarray}
	&&a(t)\!=\!(NnM_0^2/2)e^{-2\gamma\lfloor t/T\rfloor T},~\mathbf{R}\!=\!\text{diag}\{R_{1},\cdots,R_{N}\},\nonumber\\
	&&b(t)\!\!=\!\frac{\varepsilon\nu M_0\sqrt{(1\!-\!\beta)Nn}}{2\alpha\lambda_N} \!e^{-\gamma(\lfloor t/T\rfloor+1) T},~\Theta\!\!=\!\mathbf{F}(\boldsymbol{x})\!-\!\mathbf{F}(\boldsymbol{x}^*).\nonumber
\end{eqnarray}
\textbf{Step 1.} We prove that the following conclusion.
\begin{eqnarray}\label{ine1a}
	&&\|\boldsymbol{e}(t)\|\leq b(k_1T),~\forall~t\in(k_1T,t') \!\Rightarrow\!V(\overline{\boldsymbol{x}})\leq a(t)\nonumber\\
	&&~\forall~t\in(k_{1}T,t'],~\forall~t'\in\left(k_1T,(k_1+1)T\right].
\end{eqnarray}
Follow the update dynamic \eqref{algorithm} that
\begin{eqnarray}
	\dot{\overline{\boldsymbol{x}}}_{1}(t)\!&=&\!-\!(\varPhi^T_{1}\otimes I_{n})\mathbf{R}\Theta,\nonumber\\
	\dot{\overline{\boldsymbol{x}}}_{2}(t)\!&=&\!-\!\alpha(\varPhi^T_{2}L_{\mathcal{G}}\varPhi_{2}\!\otimes\! I_{n})\overline{\boldsymbol{x}}_{2}(t)\!-\!(\varPhi^T_{2}\!\otimes\!I_{n})\mathbf{R}\Theta\label{Zb}\nonumber\\
	&&~~~\!+\!\alpha(\varPhi^T_{2}L_{\mathcal{G}}\varPhi_{2}\otimes I_{n})\overline{\boldsymbol{e}}_{2}(t),\nonumber
\end{eqnarray}
Using $\varPhi_{2}\varPhi^T_{2}=I_{N}-\varPhi_{1}\varPhi^T_{1}$, compute $\dot{V}(\overline{\boldsymbol{x}})$ as
\vspace{-0.5em}
\begin{eqnarray} \dot{V}\!\!\leq-\widetilde{\boldsymbol{x}}^T\mathbf{R}\Theta\!-\!\alpha\lambda_{2}\|\overline{\boldsymbol{x}}_{2}\|^2\!+\!\alpha\overline{\boldsymbol{x}}^{T}_{2}[\varPhi^T_{2}L_{\mathcal{G}}\varPhi_{2}\otimes I_{n}]\overline{\boldsymbol{e}}_{2}.\label{dotV}
\end{eqnarray}
Since $\widetilde{\boldsymbol{x}}\!=\!(\varPhi_{1}\otimes I_{n})\overline{\boldsymbol{x}}_{1}\!+\!(\varPhi_{2}\otimes I_{n})\overline{\boldsymbol{x}}_{2}$=$\widetilde{\boldsymbol{x}}_{1}+\widetilde{\boldsymbol{x}}_{2}$, it yields
\vspace{-0.5em}
\begin{eqnarray}
	-\widetilde{\boldsymbol{x}}^T\mathbf{R}\Theta
	&&=-\widetilde{\boldsymbol{x}}_{1}^T\mathbf{R}[\mathbf{F}(\widetilde{\boldsymbol{x}}_{1}+\widetilde{\boldsymbol{x}}_{2}+\boldsymbol{x}^*)-\mathbf{F}(\widetilde{\boldsymbol{x}}_{1}+\boldsymbol{x}^*)]\nonumber\\
	&&~~-\widetilde{\boldsymbol{x}}_{2}^T\mathbf{R}[\mathbf{F}(\widetilde{\boldsymbol{x}}_{1}+\widetilde{\boldsymbol{x}}_{2}+\boldsymbol{x}^*)-\mathbf{F}(\widetilde{\boldsymbol{x}}_{1}+\boldsymbol{x}^*)]\nonumber\\
	&&~~-\widetilde{\boldsymbol{x}}_{1}^T\mathbf{R}[\mathbf{F}(\widetilde{\boldsymbol{x}}_{1}+\boldsymbol{x}^*)-\mathbf{F}(\boldsymbol{x}^*)]\nonumber\\
	&&~~-\widetilde{\boldsymbol{x}}_{2}^T\mathbf{R}[\mathbf{F}(\widetilde{\boldsymbol{x}}_{1}+\boldsymbol{x}^*)-\mathbf{F}(\boldsymbol{x}^*)].\label{11}
\end{eqnarray}
Since $\mathbf{F}(1_{N}\otimes x)=F(x)$ for any $x\in{\mathbb{R}^{n}}$ and $F(x)$ is strong monotone, the third term of \eqref{11} is written as
\vspace{-0.5em}
\begin{eqnarray}
	\widetilde{\boldsymbol{x}}_{1}^T\mathbf{R}[\mathbf{F}(\widetilde{\boldsymbol{x}}_{1}\!+\!\boldsymbol{x}^*)\!\!-\!\mathbf{F}(\boldsymbol{x}^*)]&=&\varepsilon\overline{\boldsymbol{x}}_{1}^T/(N)^{1/2}[F(\overline{\boldsymbol{x}}_{1}/(N)^{1/2}\nonumber\\
&&\!\!\!\!\!\!\!\!+x^*)\!\!-\!F(x^*)]\!\!\geqslant\!\! \varepsilon\mu/N\|\overline{\boldsymbol{x}}_{1}\|^2,\label{1}
\end{eqnarray}
where $(1^T_{N}\otimes I_{n})\mathbf{R}=\varepsilon I_{n}$ is utilized. Recalling $\|\mathbf{R}\|=\varepsilon\|\varPhi_{2}\|=\varepsilon$, it follows  from Lemma 2 that
\vspace{-0.5em}
\begin{eqnarray}
\widetilde{\boldsymbol{x}}_{2}^T\mathbf{R}[\mathbf{F}(\widetilde{\boldsymbol{x}}_{1}+\boldsymbol{x}^*)-\mathbf{F}(\boldsymbol{x}^*)]\!\leq\!  \varepsilon\theta/(N)^{1/2}\|\overline{\boldsymbol{x}}_{1}\|\|\overline{\boldsymbol{x}}_{2}\|.\label{2}
%	&&=\widetilde{\boldsymbol{x}}_{2}^T\mathbf{R}[\mathbf{F}(1_{N}\otimes(\frac{\overline{\boldsymbol{x}}_{1}^T}{\sqrt{N}}+x^*))\!-\!\mathbf{F}(1_{N}\otimes x^*))]\nonumber\\
	%&&\leq\!\varepsilon\|\overline{\boldsymbol{x}}_{2}\|\|F(\frac{\overline{\boldsymbol{x}}_{1}^T}{\sqrt{N}}\!+\!x^*)\!-\!F(x^*)\|
\end{eqnarray}
Similarly, we further obtain
\vspace{-0.5em}
\begin{eqnarray}
	&&\!-\widetilde{\boldsymbol{x}}_{1}^T\mathbf{R}\![\mathbf{F}(\widetilde{\boldsymbol{x}}_{1}\!+\!\widetilde{\boldsymbol{x}}_{2}\!+\!\boldsymbol{x}^*)\!\!-\!\mathbf{F}(\widetilde{\boldsymbol{x}}_{1}\!\!+\!\boldsymbol{x}^*)\!]\!\leq\!\varepsilon\theta/(N)^{1/2}\!\|\overline{\boldsymbol{x}}_{1}\|\!\|\!\overline{\boldsymbol{x}}_{2}\!\|,\nonumber\\
	&&	-\widetilde{\boldsymbol{x}}_{2}^T\mathbf{R}[\mathbf{F}(\widetilde{\boldsymbol{x}}_{1}\!+\!\widetilde{\boldsymbol{x}}_{2}\!+\!\boldsymbol{x}^*)\!-\!\mathbf{F}(\widetilde{\boldsymbol{x}}_{1}\!+\!\boldsymbol{x}^*)]\leq\!\varepsilon\theta\|\overline{\boldsymbol{x}}_{2}\|^2.\label{4}
\end{eqnarray}
Summing up both side of \eqref{1}-\eqref{4}, we have
\vspace{-0.5em}
\begin{eqnarray}
	-\widetilde{\boldsymbol{x}}^T\mathbf{R}\Theta\!\!\leq\! 2\varepsilon\theta/(N)^{1/2}\|\overline{\boldsymbol{x}}_{1}\|\|\overline{\boldsymbol{x}}_{2}\|\!+\!\varepsilon\theta\|\overline{\boldsymbol{x}}_{2}\|^2\!\!-\!\varepsilon\mu/N\|\overline{\boldsymbol{x}}_{1}\|^2.\nonumber
\end{eqnarray}
Then the derivative of $V$ in \eqref{dotV} is rewritten as follows,
\vspace{-0.5em}
\begin{eqnarray}
	\dot{V}&&\leq 2\varepsilon\theta/(N)^{1/2}\|\overline{\boldsymbol{x}}_{1}\|\|\overline{\boldsymbol{x}}_{2}\|+\varepsilon\theta\|\overline{\boldsymbol{x}}_{2}\|^2-\varepsilon\mu/N\|\overline{\boldsymbol{x}}_{1}\|^2\nonumber\\
	&&~~\!-\!\alpha\lambda_{2}\|\overline{\boldsymbol{x}}_{2}\|^2\!+\!\alpha\overline{\boldsymbol{x}}^{T}_{2}(\varPhi^T_{2}L_{\mathcal{G}}\varPhi_{2}\otimes I_{n})\overline{\boldsymbol{e}}_{2}\nonumber\\
%	&&\!\leq\! -\!\varepsilon[\begin{matrix}
%		\|\overline{\boldsymbol{x}}_{1}\| & \|\overline{\boldsymbol{x}}_{2}\|
%	\end{matrix}]
%	A_{\alpha/\varepsilon}
%	\bigg{[}\begin{matrix}
%		\|\overline{\boldsymbol{x}}_{1}\| \nonumber\\
%		\|\overline{\boldsymbol{x}}_{2}\|
%	\end{matrix}
%	\bigg{]}\!+\!\alpha\overline{\boldsymbol{x}}^T_{2}(\varPhi^T_{2}L_{\mathcal{G}}\varPhi_{2}\otimes I_{n})\overline{\boldsymbol{e}}_{2}, \nonumber\\
	&&{\leq}-\varepsilon\nu V \!+\!\alpha\overline{\boldsymbol{x}}^T_{2}(\varPhi^T_{2}L_{\mathcal{G}}\varPhi_{2}\otimes I_{n})\overline{\boldsymbol{e}}_{2},\label{dotVV}
\end{eqnarray}
where the second inequality holds using the fact  $\frac{\alpha}{\varepsilon}>\frac{1}{\lambda_{2}}(\frac{\theta^2}{\mu}+\theta)$.
Since $\varPhi_{2}\varPhi^T_{2}=I_{N}-\varPhi_{1}\varPhi^T_{1}$ and $1^T_{N}L=0^T_{N}$, the second term of \eqref{dotVV} is expressed as follows,
\vspace{-0.8em}
$$\alpha\overline{\boldsymbol{x}}^T_{2}(\varPhi^T_{2}L_{\mathcal{G}}\varPhi_{2}\otimes I_{n})\overline{\boldsymbol{e}}_{2}\!=\!\alpha\widetilde{\boldsymbol{x}}^T(L_{\mathcal{G}}\otimes I_{n}){\boldsymbol{e}}.$$ Thus, \eqref{dotVV} is equivalent to
\vspace{-1em}
\begin{eqnarray}
	\dot{V}\leq -\varepsilon\nu V+\alpha\widetilde{\boldsymbol{x}}^T(L_{\mathcal{G}}\otimes I_{n}){\boldsymbol{e}}.\label{dotV2}
\end{eqnarray}
It implies that
\vspace{-0.5em}
\begin{eqnarray}
	\dot{V}\leq -\varepsilon\nu V/2-\left (\varepsilon\nu \|\widetilde{\boldsymbol{x}}\|^2/4-\alpha\|\widetilde{\boldsymbol{x}}\|\|L_{\mathcal{G}}\|\|{\boldsymbol{e}}\|\right). \label{dotV1}
\end{eqnarray}
Use $\alpha\|\widetilde{\boldsymbol{x}}\|\|L_{\mathcal{G}}\|\|{\boldsymbol{e}}\|\!\leq\! \varepsilon\nu\|\widetilde{\boldsymbol{x}}\|^{2}/4\!+\!\alpha^2\|L_{\mathcal{G}}\|^2\|{\boldsymbol{e}}\|^2/(\varepsilon\nu)$ such that
$\dot{V}\!\leq\!-\! \varepsilon\nu V/2\!+\! \alpha^2\|L_{\mathcal{G}}\|^2\|{\boldsymbol{e}}\|^2/(\varepsilon\nu)$. Select $\beta\!\in\!(0,1)$ such that
$\dot{V}\leq -\varepsilon\beta \nu V/2-\varepsilon(1-\beta)\nu V/2+ \alpha^2\|L_{\mathcal{G}}\|^2\|{\boldsymbol{e}}\|^2/(\varepsilon\nu),$ and for $t\!\in\!(k_1T,(k_1+1)T),$
\vspace{-1em}
\begin{eqnarray}
	V(\overline{\boldsymbol{x}})\!\geqslant\!\! a(\!(k_1\!+\!1)T)\!>\!a(t),\|{\boldsymbol{e}}(t)\|\!<\! b(t)\! \Rightarrow \!\!\dot{V}\!\!\leq\! \frac{-\varepsilon\nu V}{2}. \label{228}
\end{eqnarray}
\vspace{-1pt}%
Since $a(t)\equiv a(k_1T),~t\in[k_1T,(k_1+1)T)$, it follows from \eqref{228} that \eqref{ine1a} holds for $t'\in(kT,(k_1+1)T)$. We then consider the situation on $t'=(k_1+1)T$. Assume that $V(\overline{\boldsymbol{x}})>a((k_1+1)T)$ for $t\in(k_{1}T,(k_{1}+1)T]$. Based on \eqref{dotV1}, there is $\dot{V}(\overline{\boldsymbol{x}})\leq -\varepsilon\nu V/2$ . Hence, $\mathcal{A}=\{\overline{\boldsymbol{x}}|V(\overline{\boldsymbol{x}})\leq a((k_1+1)T)\}$ is an invariant set. Note that ${V}(\overline{\boldsymbol{x}})\!\leq\! a(k_1T)$ when $t=k_1T$, then from \eqref{228}, $\overline{\boldsymbol{x}}(t)$ enters into the set $\mathcal{A}$ not later than the time $t\!=\!(k_1\!+\!1)T$. Thus, \eqref{ine1a} holds for $t=(k_1+1)T$.

\textbf{Step 2.} We prove the following conclusion
\vspace{-0.5em}
\begin{eqnarray}\label{ine2}
	&&V(\overline{\boldsymbol{x}})\!\leq\! a(k_1T),~\forall~t\in (k_1T, t']\Rightarrow\|\boldsymbol{e}(t)\|\!<\! b(k_1T), \nonumber \\
	&&\forall~t\!\in\!(k_1T,t'],~\forall t' \!\in\![k_1T,(k_1+1)T).\label{step2}
\end{eqnarray}
From \eqref{algorithm},
$\dot{\boldsymbol{e}}(t)\!=\!-\!\alpha\mathbf{L}\widetilde{\boldsymbol{x}}(t)\!+\!\alpha\mathbf{L}{\boldsymbol{e}}(t)\!-\!\mathbf{R}\Theta.$
It implies that
\vspace{-0.5em}
\begin{eqnarray}
	{\boldsymbol{e}}(t)\!\!&&=\!\! \e^{\alpha\mathbf{L}(t\!-\!k_1T)}\!\!{\boldsymbol{e}}(k_1T)\!\!-\!\!\!\int^{t}_{k_1T}\!\!\!\!\!\e^{\alpha\mathbf{L}(t-\tau)}\!(\!\alpha\mathbf{L}\widetilde{\boldsymbol{x}}(\tau)\!\!+\!\mathbf{R}\Theta\!)\d\tau . \label{integra}	
\end{eqnarray}
Taking the Euclidean norm of both side of \eqref{integra}, it yields
\vspace{-0.5em}
\begin{eqnarray}
	\|{\boldsymbol{e}}(t)\|\!&\leq\! \|\e^{\alpha\mathbf{L}(t-k_1T)}\|\|{\boldsymbol{e}}(k_1T)\|\!+\!\int^{t}_{k_1T}\!\|\e^{\alpha\mathbf{L}(t-\tau)}\!\!\|\nonumber \\
	&~~(\|\alpha\mathbf{L}\widetilde{\boldsymbol{x}}(\tau)\|\!+\!\|\mathbf{R}\Theta\| )\d\tau, ~k_1T<t\leq t'.\label{eeeee}
\end{eqnarray}
For any $t\in(k_1T,t']$, the first term of~\eqref{eeeee} satisfies
\vspace{-0.5em}
\begin{eqnarray}
	\|\e^{\alpha\mathbf{L}(t-k_1T)}\!\|\!\|{\boldsymbol{e}}(k_1T)\|\!\leq\!\e^{\alpha\lambda_{N}T}\|{\boldsymbol{e}}(k_1T)\|.\label{first}
\end{eqnarray}
For any $t\in(k_1T,t']$, due to $\|\widetilde{\boldsymbol{x}}(t)\|\leq NnM_0^2e^{-2\gamma t}$, then
\vspace{-0.5em}
\begin{eqnarray}
	&&~~\int^{t}_{k_1T}\!\|\e^{\alpha\mathbf{L}(t-\tau)}\| \|\alpha\mathbf{L}\widetilde{\boldsymbol{x}}(\tau)\|\d\tau\nonumber\\
	%&&\leq \alpha\lambda_{N}\int^{t}_{k_1T}\!e^{\alpha\lambda_{N}(t-\tau)}\d\tau \int^{(k_1+1)T}_{k_1T} \|\widetilde{\boldsymbol{x}}(\tau)\|\d\tau\nonumber\\
	&&\leq \alpha\lambda_{N}\sqrt{Nn}M_{0} e^{\alpha\lambda_{N}t}\int^{t}_{k_1T}\!\!e^{-\alpha\lambda_{N}\tau}\d\tau \int^{(k\!+\!1)T}_{k_1T} e^{-\frac{\varepsilon\nu\beta}{4}\tau}\d\tau\nonumber\\
%	&&\leq\! e^{\!\alpha\lambda_{N}t}(\!e^{\!-\!\alpha\lambda_{N}k_1T}\!-\!e^{\!-\!\alpha\lambda_{N}t}) \frac{4\sqrt{Nn}M_{0}}{\varepsilon\nu\beta}[\!e^{\!-\frac{\nu\varepsilon\beta}{4}\!k_1T}\!-\!e^{\!-\!\frac{\varepsilon\nu\beta}{4}\!(k_1+1)T}\!],\nonumber\\
	&&\leq \frac{4\sqrt{Nn}M_{0}}{\varepsilon\nu\beta}(e^{\alpha\lambda_{N}T}\!-\!1) (e^{\frac{\varepsilon\nu\beta}{4}T}\!-\!1)e^{-\frac{\varepsilon\nu\beta}{4}(k\!+\!1)T}.\label{second}	
\end{eqnarray}
 For any $t\in(k_1T,t']$, the third term of~\eqref{eeeee} satisfies
\vspace{-0.5em}
\begin{eqnarray}
	&&~~\int^{t}_{k_1T}\!\|\e^{\alpha\mathbf{L}(t-\tau)}\| \|\mathbf{R}\Theta\|\d\tau\nonumber\\
%	&&\leq \frac{1}{\alpha\lambda_{N}}(\e^{\alpha\lambda_{N}T}-1)\int^{(k_1+1)T}_{k_1T}\varepsilon\|\mathbf{F}(\boldsymbol{x})-\mathbf{F}(\boldsymbol{x}^*)\|\d\tau,\nonumber\\
%	&&\leq \frac{1}{\alpha\lambda_{N}}(\e^{\alpha\lambda_{N}T}-1)\|\int^{(k_1+1)T}_{k_1T}\varepsilon\theta_{\mathbf{F}}\|\boldsymbol{x}-\boldsymbol{x}^*\|\d\tau,\nonumber\\
	%&&\leq \frac{1}{\alpha\lambda_{N}}(\e^{\alpha\lambda_{N}T}-1)\varepsilon\theta_{F}\int^{(k_1+1)T}_{k_1T}\|\widetilde{\boldsymbol{x}}(\tau)\|\d\tau\nonumber\\	
	&&\leq (\e^{\alpha\lambda_{N}T}\!-\!1)\frac{4\sqrt{Nn}M_{0}\theta_{F}}{\alpha\lambda_{N}\beta\nu}(e^{\frac{\varepsilon\nu\beta}{4}T}\!-\!1)e^{-\frac{\varepsilon\nu\beta}{4}(k_1+1)T}.	\label{third}	
\end{eqnarray}
\vspace{-0.5pt}%
Denote $\Delta^k_{\boldsymbol{e}}=\sup_{t\in(k_1T,t']}\|\boldsymbol{e}(t)\|$. By \eqref{first}-\eqref{third},
\vspace{-0.5em}
\begin{eqnarray}
&\Delta^k_{\boldsymbol{e}}\!\leq\! \e^{\alpha\lambda_{N}T}\|{\boldsymbol{e}}(k_1T)\|\!+\!\left( \frac{\theta_{\mathbf{F}}\varepsilon}{\alpha\lambda_N}+1 \right) \frac{4\sqrt{Nn}M_{0}}{\varepsilon\nu\beta}\nonumber\\ &(e^{\alpha\lambda_{N}T}\!-\!1)(e^{\frac{\varepsilon\nu\beta}{4}T}-1)e^{-\frac{\varepsilon\nu\beta}{4} (k_1+1)T}.\label{bare2}
\end{eqnarray}
From  $\|\boldsymbol{e}(k_1T)\|\leq a_2e^{-\alpha\lambda_NT}b(k_1T)$, we have
\vspace{-0.5em}
\begin{eqnarray}
	\|e(t)\|\leq e^{\alpha\lambda_{N}T}\|{\boldsymbol{e}}(k_1T)\|\!\leq\! a_{2}b(k_1T),~t\in(k_1T,t'].
\end{eqnarray}
It follows from \eqref{a1} and \eqref{bare2} that
\vspace{-0.5em}
\begin{eqnarray}
	&&~\left( \frac{\theta_{\mathbf{F}}\varepsilon}{\alpha\lambda_N}\!+\!1 \right)\!\frac{4\sqrt{Nn}M_{0}}{\varepsilon\nu\beta}(e^{\alpha\lambda_{N}T}\!-\!1)\!(e^{\frac{\varepsilon\nu}{4}T}\!-\!1)e^{-\frac{\varepsilon\nu}{4} (k\!+\!1)T}\nonumber\\
	&&\leq a_{1}\frac{ \varepsilon\nu M_0}{2\alpha\lambda_N}\sqrt{Nn(1-\beta)} e^{-\frac{\varepsilon\nu}{4}\beta (k+1) T}.\label{40}
\end{eqnarray}
\vspace{-0.5em}
Since $a_{1}\!+\!a_{2}\!<\! 1$, \eqref{bare2}-\eqref{40} yields \eqref{step2}.

\textbf{Step 3.} We prove the following conclusion
\vspace{-0.5em}
\begin{eqnarray}\label{ine3}
	&&~\|\boldsymbol{e}(t)\|< b(t),~t\in[k_1T,(k_1+1)T),\nonumber\\
	&&\Rightarrow  \|\boldsymbol{e}((k_1+1)T)\|\leq a_2e^{-\alpha\lambda_NT}b((k_1+1)T).
\end{eqnarray}
\vspace{-0.5pt}%
Denote the left limit of $\boldsymbol{e}(t)$ as $\boldsymbol{e}^{-}(t)$. Since $\widehat{\boldsymbol{x}}(t)\!\!\equiv\!\widehat{\boldsymbol{x}}(k_1T),~ t\in[k_1T,(k_1+1)T),$
\vspace{-0.5em}
\begin{eqnarray}
	\boldsymbol{x}((k_1\!+\!1)T)\!-\!\widehat{\boldsymbol{x}}(k_1T)
%	&&\!\!=\!\lim_{t\rightarrow ((k_1+1)T)^-}\boldsymbol{x}(t)\!-\!\widehat{\boldsymbol{x}}(k_1T)\nonumber\\
	\!\!=\!\!\!\!\lim_{t\rightarrow ((k\!+\!1)T)^-}\!\!\!\!\boldsymbol{x}(t)\!\!-\!\widehat{\boldsymbol{x}}(t)\!
	\!\!=\!\boldsymbol{e}^{-}((k\!+\!1)T).\nonumber
\end{eqnarray}
\vspace{-0.5em}%
Using $s(k)\!=\!\frac{\alpha_2 \varepsilon\nu M_0}{\alpha\lambda_N}\sqrt{1\!-\!\beta} e^{-\alpha\lambda_NT\!-\!\frac{\varepsilon\beta\nu (k\!+\!1)T}{4}}$, we have
\vspace{-0.5em}
\begin{flalign}
	\begin{split}
		\left\|\frac{\boldsymbol{x}((k_1+1)T)-\widehat{\boldsymbol{x}}(k_1T)}{s(k_1+1)} \right\|
%		&=\left\|\frac{\boldsymbol{e}^{-}((k_1+1)T)}{s(k+1)}\right\|\nonumber\\
		&\!\leq\!\left\|\frac{b(k_1T)}{s(k_1+1)}\right\|\leq L,
	\end{split}&
\end{flalign}
that is, the quantizer is unsaturated at $t\!=\!(k_1\!+\!1)T.$ Then $\|\boldsymbol{e}((k_1+1)T)\|\!\leq\! \frac{\sqrt{Nn}}{2}s(k_{1}+1)\!=\!a_2e^{-\alpha\lambda_NT}b((k_1+1)T)$.

\textbf{Step 4.}
Based on Steps 1-3, we conclude the proof of Lemma 3. First, denote $\Omega=\{t\in(k_{1}T,(k_{1}+1)T)\big{|}||\boldsymbol{e}(t)||< b(k_{1}T)\}$, which is nonempty because of $\|\boldsymbol{e}(k_{1}T)\|<b(k_{1}T)$. 

Then, we show that $\sup_{t\in\Omega}t=(k_{1}+1)T$ via a contradiction argument. Assume that there exists $t'\in[k_{1}T,(k_{1}+1)T)$ such that $t'=\sup_{t\in\Omega}t,$ then $\|e(t')\|=b(k_1T)$, drawing on the fact that $e(t)$ is continuous on any $t\in[k_{1}T,(k_{1}+1)T)$. For any $t\in[k_{1}T,t')$, since $\|\boldsymbol{e}(t)\|< b(k_{1}T)$, it follows from \eqref{ine1a} in Step 1 that $V(\overline{\boldsymbol{x}})\leq a(t),t\in[k_{1}T,t']$. With \eqref{ine2} in Step 2, we further obtain $\|\boldsymbol{e}(t)\|<b(k_{1}T), \forall~t\in[k_{1}T,t']$, which contradicts to $\|e(t')\|=b(k_{1}T)$. Hence, $\sup_{t\in \Omega}t=(k_{1}+1)T$.

 To sum up, we can conclude that $\|\boldsymbol{e}(t)\|\!<\!b(k_1T)$ for any $t\in[k_{1}T,(k_{1}+1)T)$ and \eqref{l2} holds for $k=k_{1}+1$. Combining with  \eqref{ine1a} and \eqref{ine3}, we further conclude that  \eqref{l3} and \eqref{l1} hold for $k=k_{1}+1$.  Lemma 3 is verified.
%\begin{ack}                               % Place acknowledgements
%Partially supported by the Roman Senate.  % here.
%\end{ack}
\vspace{-1.5em}
\bibliographystyle{apacite}    % Include this if you use bibtex
\bibliography{mybibfile_nonlinear}           % and a bib file to produce the

\begin{thebibliography}{}

\bibitem [\protect \citeauthoryear {%
Chen%
\ \BBA {} Ji%
}{%
Chen%
\ \BBA {} Ji%
}{%
{\protect \APACyear {2020}}%
}]{%
xiaoqin}
\APACinsertmetastar {%
xiaoqin}%
\begin{APACrefauthors}%
Chen, Z.%
\BCBT {}\ \BBA {} Ji, H.%
\end{APACrefauthors}%
\unskip\
\newblock
\APACrefYearMonthDay{2020}{}{}.
\newblock
{\BBOQ}\APACrefatitle {Distributed Quantized Optimization Design of
  Continuous-Time Multiagent Systems Over Switching Graphs} {Distributed
  quantized optimization design of continuous-time multiagent systems over
  switching graphs}.{\BBCQ}
\newblock
\APACjournalVolNumPages{IEEE Trans. Syst. Man Cybern.Syst.}{}{}{}.
\PrintBackRefs{\CurrentBib}

\bibitem [\protect \citeauthoryear {%
Choi%
, Taleizadeh%
\BCBL {}\ \BBA {} Yue%
}{%
Choi%
\ \protect \BOthers {.}}{%
{\protect \APACyear {2020}}%
}]{%
economy}
\APACinsertmetastar {%
economy}%
\begin{APACrefauthors}%
Choi, T\BHBI M.%
, Taleizadeh, A\BPBI A.%
\BCBL {}\ \BBA {} Yue, X.%
\end{APACrefauthors}%
\unskip\
\newblock
\APACrefYearMonthDay{2020}{}{}.
\newblock
{\BBOQ}\APACrefatitle {Game theory applications in production research in the
  sharing and circular economy era} {Game theory applications in production
  research in the sharing and circular economy era}.{\BBCQ}
\newblock
\APACjournalVolNumPages{Int. J. Prod. Res.}{58}{1}{118--127}.
\PrintBackRefs{\CurrentBib}

\bibitem [\protect \citeauthoryear {%
De~Persis%
\ \BBA {} Grammatico%
}{%
De~Persis%
\ \BBA {} Grammatico%
}{%
{\protect \APACyear {2019}}%
}]{%
NEseeking5}
\APACinsertmetastar {%
NEseeking5}%
\begin{APACrefauthors}%
De~Persis, C.%
\BCBT {}\ \BBA {} Grammatico, S.%
\end{APACrefauthors}%
\unskip\
\newblock
\APACrefYearMonthDay{2019}{}{}.
\newblock
{\BBOQ}\APACrefatitle {Distributed averaging integral Nash equilibrium seeking
  on networks} {Distributed averaging integral nash equilibrium seeking on
  networks}.{\BBCQ}
\newblock
\APACjournalVolNumPages{Automatica}{110}{}{108548}.
\PrintBackRefs{\CurrentBib}

\bibitem [\protect \citeauthoryear {%
Gadjov%
\ \BBA {} Pavel%
}{%
Gadjov%
\ \BBA {} Pavel%
}{%
{\protect \APACyear {2018}}%
}]{%
NEseeking3}
\APACinsertmetastar {%
NEseeking3}%
\begin{APACrefauthors}%
Gadjov, D.%
\BCBT {}\ \BBA {} Pavel, L.%
\end{APACrefauthors}%
\unskip\
\newblock
\APACrefYearMonthDay{2018}{}{}.
\newblock
{\BBOQ}\APACrefatitle {A passivity-based approach to Nash equilibrium seeking
  over networks} {A passivity-based approach to nash equilibrium seeking over
  networks}.{\BBCQ}
\newblock
\APACjournalVolNumPages{IEEE Trans. Autom. Control}{64}{3}{1077--1092}.
\PrintBackRefs{\CurrentBib}

\bibitem [\protect \citeauthoryear {%
Hammerstein%
\ \BBA {} Selten%
}{%
Hammerstein%
\ \BBA {} Selten%
}{%
{\protect \APACyear {1994}}%
}]{%
biology}
\APACinsertmetastar {%
biology}%
\begin{APACrefauthors}%
Hammerstein, P.%
\BCBT {}\ \BBA {} Selten, R.%
\end{APACrefauthors}%
\unskip\
\newblock
\APACrefYearMonthDay{1994}{}{}.
\newblock
{\BBOQ}\APACrefatitle {Game theory and evolutionary biology} {Game theory and
  evolutionary biology}.{\BBCQ}
\newblock
\APACjournalVolNumPages{Handbook of game theory with economic
  applications}{2}{}{929--993}.
\PrintBackRefs{\CurrentBib}

\bibitem [\protect \citeauthoryear {%
Kajiyama%
, Hayashi%
\BCBL {}\ \BBA {} Takai%
}{%
Kajiyama%
\ \protect \BOthers {.}}{%
{\protect \APACyear {2021}}%
}]{%
dynamicquantizer4}
\APACinsertmetastar {%
dynamicquantizer4}%
\begin{APACrefauthors}%
Kajiyama, Y.%
, Hayashi, N.%
\BCBL {}\ \BBA {} Takai, S.%
\end{APACrefauthors}%
\unskip\
\newblock
\APACrefYearMonthDay{2021}{}{}.
\newblock
{\BBOQ}\APACrefatitle {Linear Convergence of Consensus-Based Quantized
  Optimization for Smooth and Strongly Convex Cost Functions} {Linear
  convergence of consensus-based quantized optimization for smooth and strongly
  convex cost functions}.{\BBCQ}
\newblock
\APACjournalVolNumPages{IEEE Trans. Autom. Control}{66}{3}{1254-1261}.
\newblock
\begin{APACrefDOI} \doi{10.1109/TAC.2020.2989281} \end{APACrefDOI}
\PrintBackRefs{\CurrentBib}

\bibitem [\protect \citeauthoryear {%
Li%
, Liu%
, Soh%
\BCBL {}\ \BBA {} Xie%
}{%
Li%
\ \protect \BOthers {.}}{%
{\protect \APACyear {2017}}%
}]{%
dynamicquantizer3}
\APACinsertmetastar {%
dynamicquantizer3}%
\begin{APACrefauthors}%
Li, H.%
, Liu, S.%
, Soh, Y\BPBI C.%
\BCBL {}\ \BBA {} Xie, L.%
\end{APACrefauthors}%
\unskip\
\newblock
\APACrefYearMonthDay{2017}{}{}.
\newblock
{\BBOQ}\APACrefatitle {Event-triggered communication and data rate constraint
  for distributed optimization of multiagent systems} {Event-triggered
  communication and data rate constraint for distributed optimization of
  multiagent systems}.{\BBCQ}
\newblock
\APACjournalVolNumPages{IEEE Trans. Syst. Man
  Cybern.Syst.}{48}{11}{1908--1919}.
\PrintBackRefs{\CurrentBib}

\bibitem [\protect \citeauthoryear {%
Liang%
, Yi%
\BCBL {}\ \BBA {} Hong%
}{%
Liang%
\ \protect \BOthers {.}}{%
{\protect \APACyear {2017}}%
}]{%
liangshu}
\APACinsertmetastar {%
liangshu}%
\begin{APACrefauthors}%
Liang, S.%
, Yi, P.%
\BCBL {}\ \BBA {} Hong, Y.%
\end{APACrefauthors}%
\unskip\
\newblock
\APACrefYearMonthDay{2017}{}{}.
\newblock
{\BBOQ}\APACrefatitle {Distributed Nash equilibrium seeking for aggregative
  games with coupled constraints} {Distributed nash equilibrium seeking for
  aggregative games with coupled constraints}.{\BBCQ}
\newblock
\APACjournalVolNumPages{Automatica}{85}{}{179--185}.
\PrintBackRefs{\CurrentBib}

\bibitem [\protect \citeauthoryear {%
Liu%
, Wu%
, Tian%
\BCBL {}\ \BBA {} Ling%
}{%
Liu%
\ \protect \BOthers {.}}{%
{\protect \APACyear {2021}}%
}]{%
liuyaohua}
\APACinsertmetastar {%
liuyaohua}%
\begin{APACrefauthors}%
Liu, Y.%
, Wu, G.%
, Tian, Z.%
\BCBL {}\ \BBA {} Ling, Q.%
\end{APACrefauthors}%
\unskip\
\newblock
\APACrefYearMonthDay{2021}{}{}.
\newblock
{\BBOQ}\APACrefatitle {DQC-ADMM: Decentralized Dynamic ADMM With Quantized and
  Censored Communications} {Dqc-admm: Decentralized dynamic admm with quantized
  and censored communications}.{\BBCQ}
\newblock
\APACjournalVolNumPages{EEE Trans. Neural Netw. Learning Syst.}{}{}{1-15}.
\newblock
\begin{APACrefDOI} \doi{10.1109/TNNLS.2021.3051638} \end{APACrefDOI}
\PrintBackRefs{\CurrentBib}

\bibitem [\protect \citeauthoryear {%
Lu%
, Jing%
\BCBL {}\ \BBA {} Wang%
}{%
Lu%
\ \protect \BOthers {.}}{%
{\protect \APACyear {2018}}%
}]{%
NEseeking4}
\APACinsertmetastar {%
NEseeking4}%
\begin{APACrefauthors}%
Lu, K.%
, Jing, G.%
\BCBL {}\ \BBA {} Wang, L.%
\end{APACrefauthors}%
\unskip\
\newblock
\APACrefYearMonthDay{2018}{}{}.
\newblock
{\BBOQ}\APACrefatitle {Distributed algorithms for searching generalized Nash
  equilibrium of noncooperative games} {Distributed algorithms for searching
  generalized nash equilibrium of noncooperative games}.{\BBCQ}
\newblock
\APACjournalVolNumPages{IEEE Trans. Cybern.}{49}{6}{2362--2371}.
\PrintBackRefs{\CurrentBib}

\bibitem [\protect \citeauthoryear {%
Ma%
, Ji%
, Sun%
\BCBL {}\ \BBA {} Feng%
}{%
Ma%
\ \protect \BOthers {.}}{%
{\protect \APACyear {2018}}%
}]{%
maji}
\APACinsertmetastar {%
maji}%
\begin{APACrefauthors}%
Ma, J.%
, Ji, H.%
, Sun, D.%
\BCBL {}\ \BBA {} Feng, G.%
\end{APACrefauthors}%
\unskip\
\newblock
\APACrefYearMonthDay{2018}{}{}.
\newblock
{\BBOQ}\APACrefatitle {An approach to quantized consensus of continuous-time
  linear multi-agent systems} {An approach to quantized consensus of
  continuous-time linear multi-agent systems}.{\BBCQ}
\newblock
\APACjournalVolNumPages{Automatica}{91}{}{98--104}.
\PrintBackRefs{\CurrentBib}

\bibitem [\protect \citeauthoryear {%
Nedic%
, Olshevsky%
, Ozdaglar%
\BCBL {}\ \BBA {} Tsitsiklis%
}{%
Nedic%
\ \protect \BOthers {.}}{%
{\protect \APACyear {2008}}%
}]{%
quantizationeffect2}
\APACinsertmetastar {%
quantizationeffect2}%
\begin{APACrefauthors}%
Nedic, A.%
, Olshevsky, A.%
, Ozdaglar, A.%
\BCBL {}\ \BBA {} Tsitsiklis, J\BPBI N.%
\end{APACrefauthors}%
\unskip\
\newblock
\APACrefYearMonthDay{2008}{}{}.
\newblock
{\BBOQ}\APACrefatitle {Distributed subgradient methods and quantization
  effects} {Distributed subgradient methods and quantization effects}.{\BBCQ}
\newblock
\BIn{} \APACrefbtitle {Proc. 47th IEEE Conf. Decision Control} {Proc. 47th ieee
  conf. decision control}\ (\BPGS\ 4177--4184).
\PrintBackRefs{\CurrentBib}

\bibitem [\protect \citeauthoryear {%
Nekouei%
, Nair%
\BCBL {}\ \BBA {} Alpcan%
}{%
Nekouei%
\ \protect \BOthers {.}}{%
{\protect \APACyear {2016}}%
}]{%
duibi}
\APACinsertmetastar {%
duibi}%
\begin{APACrefauthors}%
Nekouei, E.%
, Nair, G\BPBI N.%
\BCBL {}\ \BBA {} Alpcan, T.%
\end{APACrefauthors}%
\unskip\
\newblock
\APACrefYearMonthDay{2016}{}{}.
\newblock
{\BBOQ}\APACrefatitle {Performance analysis of gradient-based nash seeking
  algorithms under quantization} {Performance analysis of gradient-based nash
  seeking algorithms under quantization}.{\BBCQ}
\newblock
\APACjournalVolNumPages{IEEE Trans. Autom. Control}{61}{12}{3771--3783}.
\PrintBackRefs{\CurrentBib}

\bibitem [\protect \citeauthoryear {%
Rabbat%
\ \BBA {} Nowak%
}{%
Rabbat%
\ \BBA {} Nowak%
}{%
{\protect \APACyear {2005}}%
}]{%
quantizationeffect1}
\APACinsertmetastar {%
quantizationeffect1}%
\begin{APACrefauthors}%
Rabbat, M\BPBI G.%
\BCBT {}\ \BBA {} Nowak, R\BPBI D.%
\end{APACrefauthors}%
\unskip\
\newblock
\APACrefYearMonthDay{2005}{}{}.
\newblock
{\BBOQ}\APACrefatitle {Quantized incremental algorithms for distributed
  optimization} {Quantized incremental algorithms for distributed
  optimization}.{\BBCQ}
\newblock
\APACjournalVolNumPages{IEEE J. Sel. Areas Commun.}{23}{4}{798--808}.
\PrintBackRefs{\CurrentBib}

\bibitem [\protect \citeauthoryear {%
Salehisadaghiani%
\ \BBA {} Pavel%
}{%
Salehisadaghiani%
\ \BBA {} Pavel%
}{%
{\protect \APACyear {2016}}%
}]{%
NEseeking1}
\APACinsertmetastar {%
NEseeking1}%
\begin{APACrefauthors}%
Salehisadaghiani, F.%
\BCBT {}\ \BBA {} Pavel, L.%
\end{APACrefauthors}%
\unskip\
\newblock
\APACrefYearMonthDay{2016}{}{}.
\newblock
{\BBOQ}\APACrefatitle {Distributed Nash equilibrium seeking: A gossip-based
  algorithm} {Distributed nash equilibrium seeking: A gossip-based
  algorithm}.{\BBCQ}
\newblock
\APACjournalVolNumPages{Automatica}{72}{}{209--216}.
\PrintBackRefs{\CurrentBib}

\bibitem [\protect \citeauthoryear {%
Shoham%
}{%
Shoham%
}{%
{\protect \APACyear {2008}}%
}]{%
computersciences}
\APACinsertmetastar {%
computersciences}%
\begin{APACrefauthors}%
Shoham, Y.%
\end{APACrefauthors}%
\unskip\
\newblock
\APACrefYearMonthDay{2008}{}{}.
\newblock
{\BBOQ}\APACrefatitle {Computer science and game theory} {Computer science and
  game theory}.{\BBCQ}
\newblock
\APACjournalVolNumPages{Commun. ACM}{51}{8}{74--79}.
\PrintBackRefs{\CurrentBib}

\bibitem [\protect \citeauthoryear {%
Ye%
\ \BBA {} Hu%
}{%
Ye%
\ \BBA {} Hu%
}{%
{\protect \APACyear {2017}}%
}]{%
NEseeking2}
\APACinsertmetastar {%
NEseeking2}%
\begin{APACrefauthors}%
Ye, M.%
\BCBT {}\ \BBA {} Hu, G.%
\end{APACrefauthors}%
\unskip\
\newblock
\APACrefYearMonthDay{2017}{}{}.
\newblock
{\BBOQ}\APACrefatitle {Distributed Nash equilibrium seeking by a consensus
  based approach} {Distributed nash equilibrium seeking by a consensus based
  approach}.{\BBCQ}
\newblock
\APACjournalVolNumPages{IEEE Trans. Autom. Control}{62}{9}{4811--4818}.
\PrintBackRefs{\CurrentBib}

\bibitem [\protect \citeauthoryear {%
Yi%
\ \BBA {} Hong%
}{%
Yi%
\ \BBA {} Hong%
}{%
{\protect \APACyear {2014}}%
}]{%
dynamicquantizer1}
\APACinsertmetastar {%
dynamicquantizer1}%
\begin{APACrefauthors}%
Yi, P.%
\BCBT {}\ \BBA {} Hong, Y.%
\end{APACrefauthors}%
\unskip\
\newblock
\APACrefYearMonthDay{2014}{}{}.
\newblock
{\BBOQ}\APACrefatitle {Quantized subgradient algorithm and data-rate analysis
  for distributed optimization} {Quantized subgradient algorithm and data-rate
  analysis for distributed optimization}.{\BBCQ}
\newblock
\APACjournalVolNumPages{IEEE Trans. Control Netw. Syst.}{1}{4}{380--392}.
\PrintBackRefs{\CurrentBib}

\bibitem [\protect \citeauthoryear {%
You%
\ \BBA {} Xie%
}{%
You%
\ \BBA {} Xie%
}{%
{\protect \APACyear {2011}}%
}]{%
youkeyou}
\APACinsertmetastar {%
youkeyou}%
\begin{APACrefauthors}%
You, K.%
\BCBT {}\ \BBA {} Xie, L.%
\end{APACrefauthors}%
\unskip\
\newblock
\APACrefYearMonthDay{2011}{}{}.
\newblock
{\BBOQ}\APACrefatitle {Network topology and communication data rate for
  consensusability of discrete-time multi-agent systems} {Network topology and
  communication data rate for consensusability of discrete-time multi-agent
  systems}.{\BBCQ}
\newblock
\APACjournalVolNumPages{IEEE Trans. Autom. Control}{56}{10}{2262--2275}.
\PrintBackRefs{\CurrentBib}

\bibitem [\protect \citeauthoryear {%
Yuan%
, Xu%
, Zhao%
\BCBL {}\ \BBA {} Rong%
}{%
Yuan%
\ \protect \BOthers {.}}{%
{\protect \APACyear {2012}}%
}]{%
yuandeming}
\APACinsertmetastar {%
yuandeming}%
\begin{APACrefauthors}%
Yuan, D.%
, Xu, S.%
, Zhao, H.%
\BCBL {}\ \BBA {} Rong, L.%
\end{APACrefauthors}%
\unskip\
\newblock
\APACrefYearMonthDay{2012}{}{}.
\newblock
{\BBOQ}\APACrefatitle {Distributed dual averaging method for multi-agent
  optimization with quantized communication} {Distributed dual averaging method
  for multi-agent optimization with quantized communication}.{\BBCQ}
\newblock
\APACjournalVolNumPages{Syst. Control Lett.}{61}{11}{1053--1061}.
\PrintBackRefs{\CurrentBib}

\bibitem [\protect \citeauthoryear {%
Zeng%
, Chen%
, Liang%
\BCBL {}\ \BBA {} Hong%
}{%
Zeng%
\ \protect \BOthers {.}}{%
{\protect \APACyear {2019}}%
}]{%
NEseeking6}
\APACinsertmetastar {%
NEseeking6}%
\begin{APACrefauthors}%
Zeng, X.%
, Chen, J.%
, Liang, S.%
\BCBL {}\ \BBA {} Hong, Y.%
\end{APACrefauthors}%
\unskip\
\newblock
\APACrefYearMonthDay{2019}{}{}.
\newblock
{\BBOQ}\APACrefatitle {Generalized Nash equilibrium seeking strategy for
  distributed nonsmooth multi-cluster game} {Generalized nash equilibrium
  seeking strategy for distributed nonsmooth multi-cluster game}.{\BBCQ}
\newblock
\APACjournalVolNumPages{Automatica}{103}{}{20--26}.
\PrintBackRefs{\CurrentBib}

\bibitem [\protect \citeauthoryear {%
Zhu%
, Yu%
, Wen%
\BCBL {}\ \BBA {} Chen%
}{%
Zhu%
\ \protect \BOthers {.}}{%
{\protect \APACyear {2020}}%
}]{%
yuwenwu}
\APACinsertmetastar {%
yuwenwu}%
\begin{APACrefauthors}%
Zhu, Y.%
, Yu, W.%
, Wen, G.%
\BCBL {}\ \BBA {} Chen, G.%
\end{APACrefauthors}%
\unskip\
\newblock
\APACrefYearMonthDay{2020}{}{}.
\newblock
{\BBOQ}\APACrefatitle {Distributed Nash Equilibrium Seeking in an Aggregative
  Game on a Directed Graph} {Distributed nash equilibrium seeking in an
  aggregative game on a directed graph}.{\BBCQ}
\newblock
\APACjournalVolNumPages{IEEE Trans. Autom. Control}{66}{6}{2746--2753}.
\PrintBackRefs{\CurrentBib}

\end{thebibliography}
                                 % bibliography (preferred). The

%%%%%%%%%%%%%%%%%%%%%%%%%%%%%%%%%%%%%%%%%%%%%%%%%%%%%%%%%%%%%%%%%%

\end{document}